\documentclass[twocolumn,linenumbers]{aastex63}

\nolinenumbers

\usepackage{breqn}

\graphicspath{{./}{figures/}}

\newcommand{\Msun}{\ensuremath{M_{\odot}}}

\shorttitle{}
\shortauthors{Stachie et al.}

\begin{document}

\title{Searches for Modulated $\gamma$-Ray Precursors to Compact Binary Mergers in Fermi-GBM Data}

\correspondingauthor{Cosmin Stachie}
\email{scosmin@oca.eu}

\author{Cosmin Stachie}
\affiliation{Artemis, Universit\'e C\^ote d’Azur, Observatoire de la C\^ote d’Azur, CNRS, Nice 06300, France}

\author[0000-0001-5078-9044]{Tito \surname{Dal Canton}}
\affiliation{NASA Goddard Space Flight Center, 8800 Greenbelt Rd., Greenbelt, MD 20771, USA}
\affiliation{Universit\'e Paris-Saclay, CNRS/IN2P3, IJCLab, 91405 Orsay, France}

\author[0000-0002-6870-4202]{Nelson Christensen}
\affiliation{Artemis, Universit\'e C\^ote d’Azur, Observatoire de la C\^ote d’Azur, CNRS, Nice 06300, France}

\author[0000-0002-4618-1674]{Marie-Anne Bizouard}
\affiliation{Artemis, Universit\'e C\^ote d’Azur, Observatoire de la C\^ote d’Azur, CNRS, Nice 06300, France}

\author{Michael Briggs}
\affiliation{Department of Space Science, University of Alabama in Huntsville, Huntsville, AL 35899, USA}

\author[0000-0002-2942-3379]{Eric Burns}
\affiliation{Department of Physics \& Astronomy, Louisiana State University, Baton Rouge, LA 70803, USA}

\author{Jordan Camp}
\affiliation{NASA Goddard Space Flight Center, 8800 Greenbelt Rd., Greenbelt, MD 20771, USA}

\author[0000-0002-8262-2924]{Michael Coughlin}
\affiliation{School of Physics and Astronomy, University of Minnesota, Minneapolis, Minnesota 55455, USA}

\begin{abstract}
\nolinenumbers
GW170817 is the only gravitational-wave (GW) event, for which a confirmed $\gamma$-ray counterpart, GRB 170817A, has been detected. Here we present a method to search for another type of $\gamma$-ray signal, a $\gamma$-ray burst precursor, associated with a compact binary merger. If emitted shortly before the coalescence, a high-energy electromagnetic (EM) flash travels through a highly dynamical and relativistic environment, created by the two compact objects orbiting each other. Thus, the EM signal arriving at an Earth observer could present a somewhat predictable time-dependent modulation. We describe a targeted search method for lightcurves exhibiting such a modulation, parameterized by the observer-frame component masses and binary merger time, using Fermi-GBM data. The sensitivity of the method is assessed based on simulated signals added to GBM data. 
The method is then applied to a selection of potentially interesting compact binary mergers detected during the second (O2) and third (O3) observing runs of Advanced LIGO and Advanced Virgo. We find no significant modulated $\gamma$-ray precursor signal associated with any of the considered events.
\end{abstract}


\section{Introduction} 
\label{sec:intro}

Multimessenger astronomy started with the detection of the core-collapse supernova 1987A in both the electromagnetic (EM) and neutrino channels~\citep{doi:10.1146/annurev.aa.27.090189.003213}. Since then, only one other unambiguous example of astrophysical event heralded by different messengers occurred: the simultaneous detection of the gravitational waves (GWs) from the binary neutron star (BNS) coalescence GW170817~\citep{PhysRevLett.119.161101}, and of several EM counterpart signals: the high energy photons of GRB 170817A \citep{Goldstein_2017, Savchenko_2017}, the ultra-violet, optical and infrared radiation of the kilonova AT 2017gfo \citep{Coulter1556, 2017Natur.551...75S} and the X-ray, optical and radio afterglow of the $\gamma$-ray burst \citep{Lamb_2019, refId0_Avanzo}.
Additionally, there is a convincing claim for the coincident detection of the high energy neutrino IceCube-170922A and the multi-wavelength EM radiation coming from the $\gamma$-ray blazar TXS 0506+056 \citep{IceCube:2018dnn}.

In the observable Universe, compact objects such as neutron stars and black holes are often found in pairs, forming binaries.
During the inspiral, they lose angular momentum and binding energy, by emission of GWs \citep{1982ApJ...253..908T}.
This  implies a narrowing of the distance separating the binary components, leading in some cases to a merger in less than a Hubble time~\citep{1991ApJ...380L..17P}.
Both the frequency and the amplitude of the GWs increase as the merger approaches; during the seconds prior to the merger, the frequency of the GWs sweeps from tens of $\mbox{Hz}$ to above a $\mbox{kHz}$.
If the binary is close enough to us, the GW strain is above the sensitivity threshold of and thereby detectable by GW detectors such as Advanced LIGO~\citep{2015CQGra..32g4001L} and Advanced Virgo~\citep{TheVirgo:2014hva}. 

One of the EM counterparts to compact binary mergers is the flash of $\gamma$- and X-rays from the short $\gamma$-ray burst~\citep{doi:10.1146/annurev-astro-081913-035926, Kochanek:1993mw}, lasting less than $2\,\text{s}$~\citep{1993ApJ...413L.101K} and possessing an isotropic-equivalent energy up to $10^{51}\,\text{erg}$~\citep{DAVANZO201573}.
This short, high-energy EM signal is followed by a longer-lasting, less energetic radiation, the $\gamma$-ray burst afterglow \citep{2005Natur.437..845F}
covering a broad range of the EM spectrum, from X-ray~\citep{Vietri:1997fca}, through optical~\citep{Meszaros:1996sv} to radio~\citep{Paczynski:1993gz}.
The kilonova comprises the ultraviolet, optical and infrared radiation associated with the radioactive decay of heavy elements~\citep{1974ApJ...192L.145L, Li:1998bw, 10.1111/j.1365-2966.2010.16864.x, Kasen:2017sxr}; it has a quasi-thermal~\citep{Kasen_2019} and quasi-isotropic emission~\citep{2020ApJ...897..150D}, very different from the $\gamma$-ray burst afterglows which display large anisotropies~\citep{10.1111/j.1365-2966.2010.17616.x}, and consequently require nearly-aligned observer positions in order to be detectable.

While the association between the merger of compact objects and some $\gamma$-ray burst related messengers, such as the $\gamma$-ray prompt emission, the afterglow and the kilonova, has been unambiguously highlighted by GW170817, the presence of other kinds of EM emission is still debated.
One such example is the precursor activity to short $\gamma$-ray bursts.
\cite{Troja_2010} claim that up to $10\%$ of short $\gamma$-ray bursts, detected by Swift~\citep{Barthelmy:2005hs}, possess EM precursors, lasting less than $1\,\text{s}$ and whose starting emission might go back about $100\,\text{s}$ before the main $\gamma$-ray burst.
The existence of such precursor signals is also supported by \cite{Zhu2015}.
The investigation by~\cite{Li:2020nhs} concludes that the precursors have shorter duration than the prompt $\gamma$-ray emission, but seem to be produced by similar central engine activity.
16 precursors to short $\gamma$-ray bursts in Fermi-GBM data have been found by ~\cite{Wang:2020vvr}, who infer comparable duration for the main and precursor emissions, and possible fits with blackbody, non-thermal cutoff power law models.
Precursor emission is also motivated on a theoretical basis.
Two popular models are the resonant shattering of the crust of a neutron star during the inspiral \citep{Neill:2021lat, PhysRevLett.108.011102} and the ``black hole battery'' model \citep{McWilliams:2011zi}.
Theoretical work has also been done to identify possible features of an EM signal emitted during the inspiral, namely a modulation induced by the orbital motion \citep{Schnittman:2017nhg}.

The prompt $\gamma$ emission, afterglows and kilonovae described above are currently expected to follow from the disruption of a neutron star and the formation of an accretion disk during the merger of the binary.
However, whether a neutron star will actually disrupt before merging with the companion strongly depends on the properties of the two objects, in particular on their masses, spins and structure \citep{PhysRevD.98.081501}.
Although BNS mergers are generally always expected to radiate the whole plethora of EM signals observed with GW170817, this is far from being guaranteed for neutron star - black hole (NSBH) mergers.
In addition, the prompt $\gamma$ emission is expected to be detectable only for very specific orientations of the binary with respect to the observer.
Hence, the idea of a premerger, precursor emission that does not require disruption and may not be strongly anisotropic becomes particularly interesting: it may be the only EM signal systematically emitted by NSBH mergers, even those involving non-spinning black holes more massive than $\approx 10\ \Msun$.

Motivated by the above consideration, and by the present ambiguity regarding the possibility of precursor emission, we propose a method to analyze archival $\gamma$-ray data in temporal proximity to GW events associated with BNS mergers.
Following the idea of \citet{Schnittman:2017nhg}, the analysis aims at detecting pulsations in the $\gamma$-ray data having the same orbital phase evolution like the GW inspiral signal, during the last several seconds before merger.
The method is an extension of a more generic search for $\gamma$-ray transients based on a likelihood approach, introduced by \citet{Blackburn:2014rqa} and further optimized in \citet{Goldstein:2016zfh} and \citet{Goldstein:2019tfz}.
We demonstrate how the existing and proposed methods recover simulated signals added to archival data, and we apply the proposed method to data around a selection of compact binary merger events detected by Advanced LIGO and Advanced Virgo.

We employ data from the $\gamma$-ray Burst Monitor (GBM) instrument on the Fermi spacecraft \citep{Meegan_2009}. GBM is an ideal instrument to study rapidly-evolving $\gamma$-ray counterparts to compact binary mergers. It is able to perform time-resolved spectroscopy of high-energy EM radiation, thanks to its twelve sodium iodide (NaI) and two bismuth germanate (BGO) scintillation detectors, covering an energy range from $8\,\text{keV}$ to $40\,\text{MeV}$~\citep{ATWOOD1994302}, and distributed around the spacecraft. It has a temporal resolution of $2\,\mu\text{s}$, suitable to study the rapid modulations we are interested in, and can discriminate between $128$ energy ranges/channels. Its field of view is only limited by the Earth, and its observing time is mostly limited by passages through the South Atlantic Anomaly, implying it can witness an astrophysical transient event 75\% of the time.

The paper is structured as follows: in Section~\ref{sec:gamma_ray_precursor}, we discuss the possible physical mechanisms responsible for the $\gamma$-ray 
precursor activity, we introduce a modulated EM
waveform model and present how it can be used to simulate data from a $\gamma$-ray detector. The statistical search
method is presented in Section~\ref{sec:statistical_method}. The sensitivity of the
search is presented in Section~\ref{sec:sensitivity}, and the results of
the search around some of the LIGO-Virgo events appear in
Section~\ref{sec:real_events}. Finally, Section~\ref{sec:conclusion} offers the conclusion.

\section{Modulated $\gamma$-ray precursors}
\label{sec:gamma_ray_precursor}
\subsection{Physical mechanisms and signal models}
While the physical mechanism responsible for such a $\gamma$-ray precursor is still in question, several possibilities have been suggested in the literature. 
\cite{PhysRevLett.111.061105} and \cite{Most:2020ami} show that magnetosphere interaction in a BNS can power EM radiation, prior to the main emission, with luminosities reaching $10^{45}\, \text{erg/s}$. Copious EM radiation, prior to the merger of a magnetized neutron star-spinning black hole binary, can also be emitted through the unipolar inductor mechanism~\citep{DOrazio:2015jcb, DOrazio:2013ngp, McWilliams:2011zi, Palenzuela:2011es}: indeed an electric circuit is formed, where the roles of battery, resistor, and electrical wires are played by the black hole, the neutron star and its magnetosphere, and the magnetic field lines; in this way, black hole rotational energy is extracted by the magnetic field and sent far away by means of powerful, collimated Poynting flux.

During the last orbits of a quasicircular inspiral, or a periastron passage in an eccentric or
hyperbolic close encounter, a neutron star can experience tidal forces that excite some of its many oscillation modes, e.g.~core, shear, crustal discontinuity modes~\citep{Lai:1993di, 10.1143/ptp/91.5.871}. This process, especially when resonant, can cause quakes and/or shattering of the neutron star crust, followed by the release of a huge amount of energy in the form of EM radiation \citep{PhysRevD.101.083002, PhysRevLett.108.011102, article_Reisenegger, Tsang_2013}.

While traditionally $\gamma$-ray burst emissions are detected by searching for an excess of photons with respect to the background~\citep{Meegan_2009, Kocevski:2018suj, 2016arXiv161007385B}, in the present work we attempt to increase the sensitivity of the pipeline by hunting for well-characterized EM waveforms. ~\cite{Schnittman:2017nhg} propose a method to calculate the EM precursor lightcurve from a NSBH, where the surface of the neutron star is uniformly shining. A similar method is proposed in~\cite{Haiman:2017szj} for the case of a supermassive black hole binary, which are future LISA~\citep{2017arXiv170200786A} signal progenitors. In the preceding papers, the lightcurve is modulated by physical processes such as relativistic beaming and gravitational lensing. There is a parallelism to the compact binary coalescence GW waveforms as the EM lightcurve could be locked in phase with the GW signal. As the merger time approaches, the amplitude and the frequency of the presumed EM signal increase. Radio precursor lightcurves to compact binaries containing at least one magnetized neutron star are also suggested in~\cite{2021MNRAS.501.3184S}.

The association of GWs and $\gamma$-ray burst precursors may represent an unique class of multimessenger events, in the coming years. Indeed, the theoretically-motivated signals presented above are expected to be visible only for nearby events. As GW detections will be nearby too, this is a huge gain for this kind of work. In addition, NSBH systems, for which the mass ratio is too high, are not expected to generate neither short $\gamma$-ray bursts nor kilonovae because the neutron star component is swallowed by the black hole companion before being disrupted by the tidal field. On the other hand, these systems might power precursors of $\gamma$-ray burst, during the inspiral phase. Moreover, lightcurve models like the one proposed in~\cite{Schnittman:2017nhg} are interesting because they give information about the potential $\gamma$-ray signal we want to detect, based on the GW detection. 

\subsection{Orbital-modulation model}

An important feature of an EM lightcurve is the evolution of the brightness with time.
In this work, given a binary inspiral, we are interested in the orbital-phase-dependence of the luminosity.
For a compact binary, the first Post-Newtonian expansion \citep{Blanchet:2013haa} term of the orbital angular frequency evolution can be written as
\begin{equation}
 \Omega(t) = \left(\frac{5}{256}\frac{1}{t_c - t}\right)^{3/8}\left(\frac{G\mathcal{M}}{c^3}\right)^{-5/8},   
 \label{eqn:orbital_frequency}
\end{equation}
where $\mathcal{M} = (m_1 m_2)^{3/5} (m_1 + m_2)^{-1.5}$, $G$ and $c$ are the observer-frame chirp mass, the gravitational constant and the speed of light in vacuum respectively.
$t_c$ and $t$ are the merger and variable time measured in the observer frame.
$m_1$ and $m_2$ are the binary component masses.
We define the orbital phase, $\Phi_{\rm orbit}(t) \in [0, 2\pi]$, with origin at $t=-30\,\text{s}$, by $\Phi_{\rm orbit}(t) = \int_{-30\,\text{s}}^{t}\Omega(x)dx \mod 2\pi$. 

\begin{figure}[tb]
    \includegraphics[width=\columnwidth]{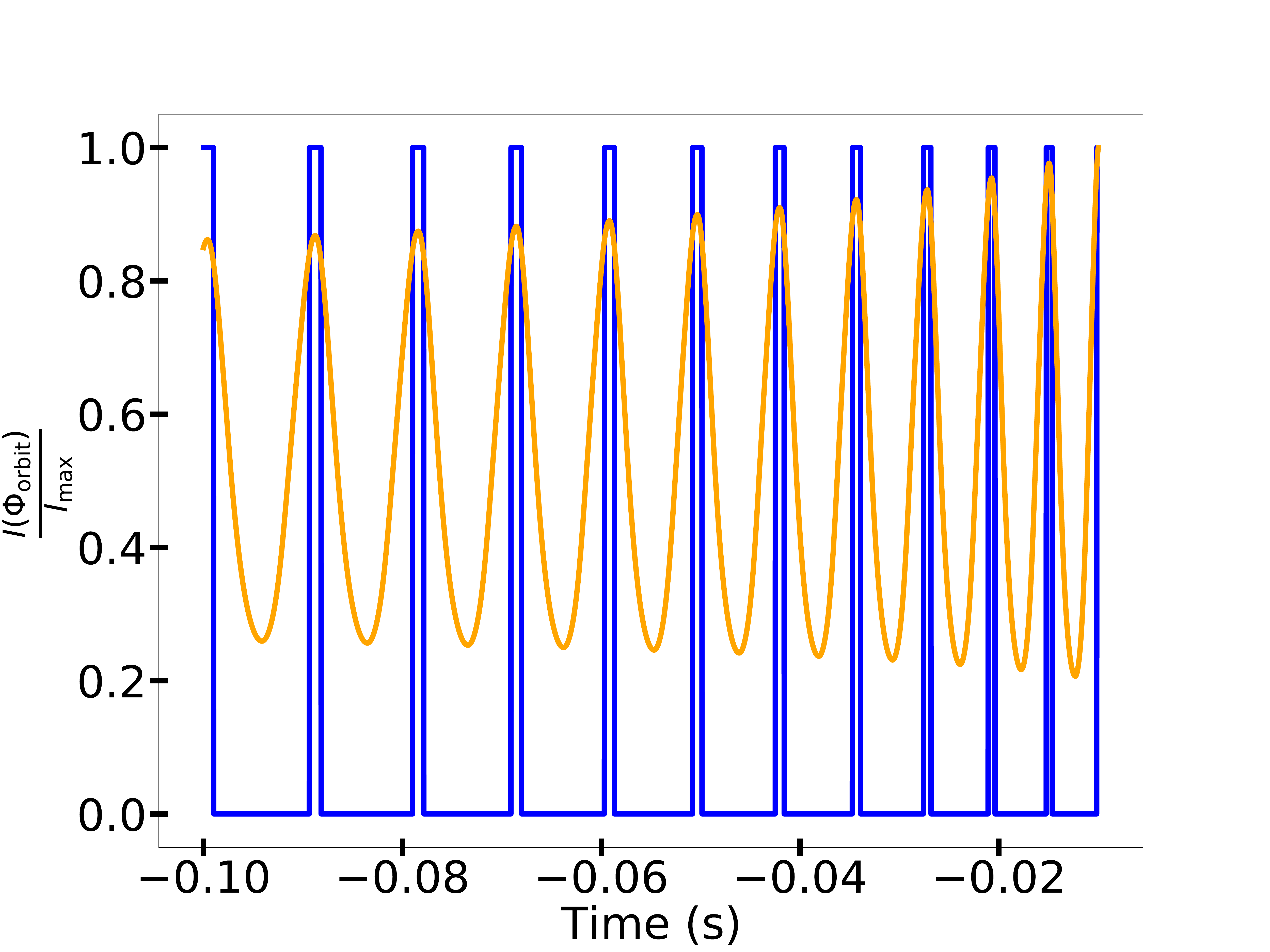}
    \caption{Normalized light intensity versus time. In blue, the \emph{simplified lightcurve} with $\theta_{\rm peak} = 270\,^\circ$ and $\theta_{\rm width} = 40\,^\circ$. 
    In orange, a lightcurve obtained by the combination of relativistic beaming and gravitational lensing effects, for which we chose an inclination angle of $45\,^\circ$ and a spectral index equal to 0. For both lightcurves, the compact objects are assumed to have masses $m_1 = 10\, M_\odot$ and $m_2 = 1.4\,M_\odot$.} 
    \label{fig:waveform_parameters}
\end{figure}

We also introduce the \emph{simplified lightcurve}, a lightcurve which, besides the chirp mass, depends on two other parameters, namely $\theta_{\rm peak}$ and $\theta_{\rm width}$.  The normalized luminosity has the following expression:
\begin{eqnarray}
   \frac{I\left(\Phi_{\rm orbit}\right)}{I_{\rm max}} = \left\{\begin{array}{ll}
        1 
       \hspace{0.2cm} \text{if } |\Phi_{\rm orbit} - \theta_{\rm peak}| \leq \theta_{\rm width}/2 
        \\
        0
    \hspace{0.2cm} \text{otherwise} \\
        \end{array}. \right.
    \label{eqn:normalized_exotic}
\end{eqnarray}
This expression corresponds to a compact binary emitting only during the orbital phase window centered at $\theta_{\rm peak}$, with width $\theta_{\rm width}$. In Figure~\ref{fig:waveform_parameters}, there is an illustration of a simplified lightcurve.
Additionally, for comparison purposes, we show a more realistic (albeit still approximate) lightcurve which attempts to explicitly model the relativistic beaming and gravitational lensing effects described in \cite{Schnittman:2017nhg}. Although we will not use this model for the analyses described later, more details of its construction are presented in Appendix~\ref{sec:realistic_lightcurve}.

Figure~\ref{fig:waveform_parameters} shows that the blue simplified lightcurve and the more realistic orange lightcurve are similar. The choice of the simplified lightcurve for the present paper is motivated by the pronounced modulation feature, limiting the spread of incoming photons in regions where the light flux is negligible. While the exacerbated modulation of the simplified lightcurve is most likely unrealistic, the preference for this model is mainly motivated by our attempt to check the validity of the search method in the case of signals with optimistically large modulation.

\subsection{Simulating a Fermi-GBM observation}

We use Fermi-GBM's Time-Tagged Event (TTE) data\footnote{\url{https://fermi.gsfc.nasa.gov/ssc/data/access/gbm/}}, which consists of a list of photons characterized by their arrival time (to microsecond precision) and energy channel. In our analysis, for simplicity and to increase the photon statistic, we coalesce the $128$ possible energy channels into $8$ main channels, larger in energy.

In order to simulate a simplified-lightcurve-like signal, one needs to convert the spectra into detector counts. To this end, one needs the detector response to radiation, characterized by the energy of its photons, the arrival direction and the amplitude of the lightcurve. This function includes two components: the response due to the direct radiation~\citep{article_Kippen} as well as the response due to the scattering from both the Earth's atmosphere and the spacecraft~\citep{Pendleton_1999}. The spectral and directional dependence was validated experimentally by ground-based calibration~\citep{2009ExA....24...47B}.
As explained in~\cite{Connaughton_2015}, based on the numerical values of the detector response taken at 41,168 grid points (accounting for 272 sky directions), the response to any arrival direction is constructed by interpolation among the three closest grid points, obtained by Delaunay triangulation. 
Regarding the radiation energetics, we make use of three spectral templates, which we refer to as: \textit{soft} (lowest energy), \textit{normal} and \textit{hard} (highest energy). For the \textit{soft} and \textit{normal} templates, we use the Band parameterized functions~\citep{1993ApJ...413..281B}, introduced in~\cite{Connaughton_2015}.
Thus the flux of photons having an energy in $[E, E + dE]$ is proportional to
\begin{widetext}
\begin{equation}
\left\{\begin{array}{ll}
        \left(\frac{E}{100\ \mathrm{keV}}\right)^{\alpha_{\rm spec}}e^{-\frac{(\alpha_{\rm spec} + 2)E}{E_{\rm peak}}} &  
       \text{if } E < \frac{(\alpha_{\rm spec} -\beta_{\rm spec})E_{\rm peak}}{\alpha_{\rm spec} + 2} 
         \\
         \\
        \left(\frac{E}{100\ \mathrm{keV}}\right)^{\beta_{\rm spec}}e^{(\beta_{\rm spec} - \alpha_{spec})}\left[\frac{(\alpha_{\rm spec} - \beta_{\rm spec})E_{\rm peak}}{(\alpha_{\rm spec} + 2)100\ \mathrm{keV}}\right]^{(\alpha_{\rm spec} - \beta_{\rm spec})} & 
     \text{otherwise}\\
        \end{array},\right.
\label{eqn:Band function}
\end{equation}
\end{widetext}
with $(\alpha_{\rm spec}, \beta_{\rm spec}, E_{\rm peak})$ equal to $(-1.9, -3.7, 70\,\text{keV})$ and $(-1, -2.3, 230\,\text{keV})$ for the \textit{soft} and \textit{normal} spectrum. With respect to the \textit{hard} spectrum, we employ the comptonized template proposed in~\cite{Goldstein:2016zfh}, such that the photon flux is proportional to 
\begin{equation}
\left(\frac{E}{E_{\rm piv}}\right)^{\alpha_{\rm spec}}\exp{\left[-\frac{(\alpha_{\rm spec} + 2) E}{E_{\rm peak}}\right]},
\end{equation}
where $E_{\rm piv}$ is a constant and $(\alpha_{\rm peak}, E_{\rm peak})$ equals $(-0.5, 1.5\,\text{MeV})$. However, the  $\gamma$-ray bursts detected to date reveal a richer set of spectra than the three templates presented here. This fact motivates our choice of considering injected lightcurves parameterized by the variable $\kappa \in [0, 2]$, such that $\kappa = 0, 1$ and $2$ correspond to the \textit{hard}, $\textit{normal}$ and \textit{soft} spectrum. For all the other values, we consider linear combinations of the three templates with the following weights:
\begin{equation}
    (w_0, w_1, w_2) = 
    \left\{\begin{array}{lr}
        (1 - \kappa, \kappa, 0) & \text{if } \kappa \in (0, 1) \\
        \\
        (0, 2 - \kappa, \kappa - 1) & \text{if }  \kappa \in (1, 2)\\
        \end{array},\right.
\label{eqn:spectral_weights}
\end{equation}
where $w_0$, $w_1$ and $w_2$ are the weights attributed to the \textit{hard}, \textit{normal} and \textit{soft} spectrum templates.

The detector output to an input flux of high-energy photons is captured by the Response Matrix. More precisely, the Response Matrix is a $14 \times 8$ array, where each row designates one of the 14 detectors and each column stands for an energy channel. Finally, each element of the Response Matrix represents the photon rate as a function of time. 
As the TTE data counts are assigned arrival times, we convert the photon rate function into a time histogram using a Poisson distribution. Moreover, for technical reasons, the data is binned. More precisely, the time is divided in intervals of size equal to $0.5\,\text{ms}$, and the photons found in the same interval/bin and belonging to the same energy channel are summed. According to the Nyquist-Shannon sampling theorem, such a binning allows the preservation of signal frequency components up to $1000\,\text{Hz}$. According to Equation~\ref{eqn:orbital_frequency}, the orbital frequency is always lower than this upper limit, for all binaries where the heavier compact object weighs more than $1.4\,M_\odot$ and for a binary evolution up to the last millisecond before the merger. Finally a simplified lightcurve injection is completely parameterized by the tuple $(t_c, m_1, m_2, f_{\rm amp}, t_{\rm start}, \Delta t_{\rm dur}, \theta_{\rm peak}, \theta_{\rm width}, \theta_{\rm ra}, \theta_{\rm dec}, \kappa)$, where the variables $f_{\rm amp}$, $t_{\rm start}$, $\Delta t_{\rm dur}$, $\theta_{\rm ra}$ and $\theta_{\rm dec}$ designate the lightcurve amplitude factor, the EM signal start time, the signal duration, the right ascension and the declination associated with the sky location. It is worth mentioning that the photon flux is obtained by the multiplication of the amplitude factor $f_{\rm amp}$ with the expression of the normalized simplified lightcurve proposed in Equation~\ref{eqn:normalized_exotic}.

\section{Statistical search method} 
\label{sec:statistical_method}

\begin{figure}
    \includegraphics[width=\columnwidth]{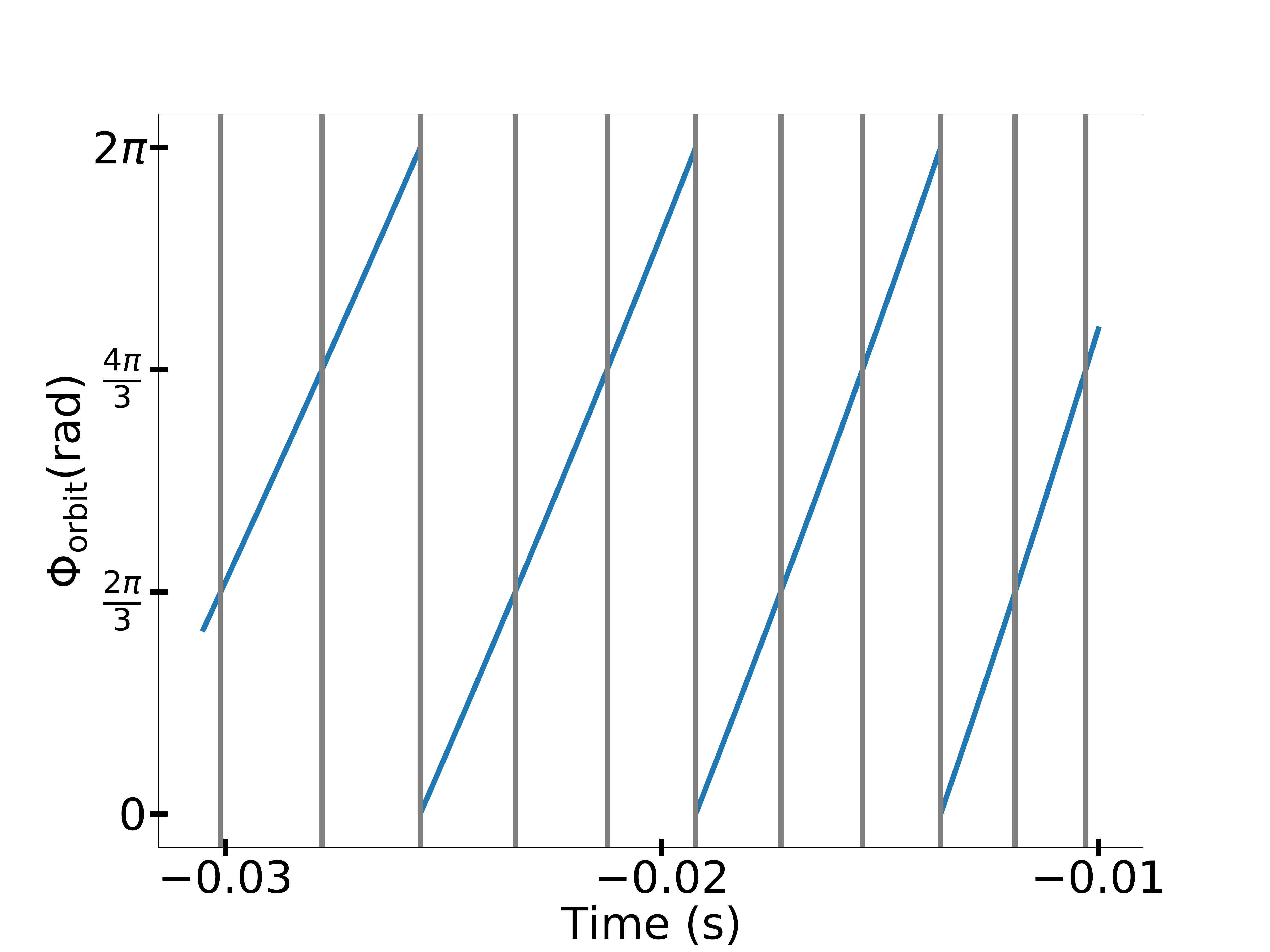}
    \includegraphics[width=\columnwidth]{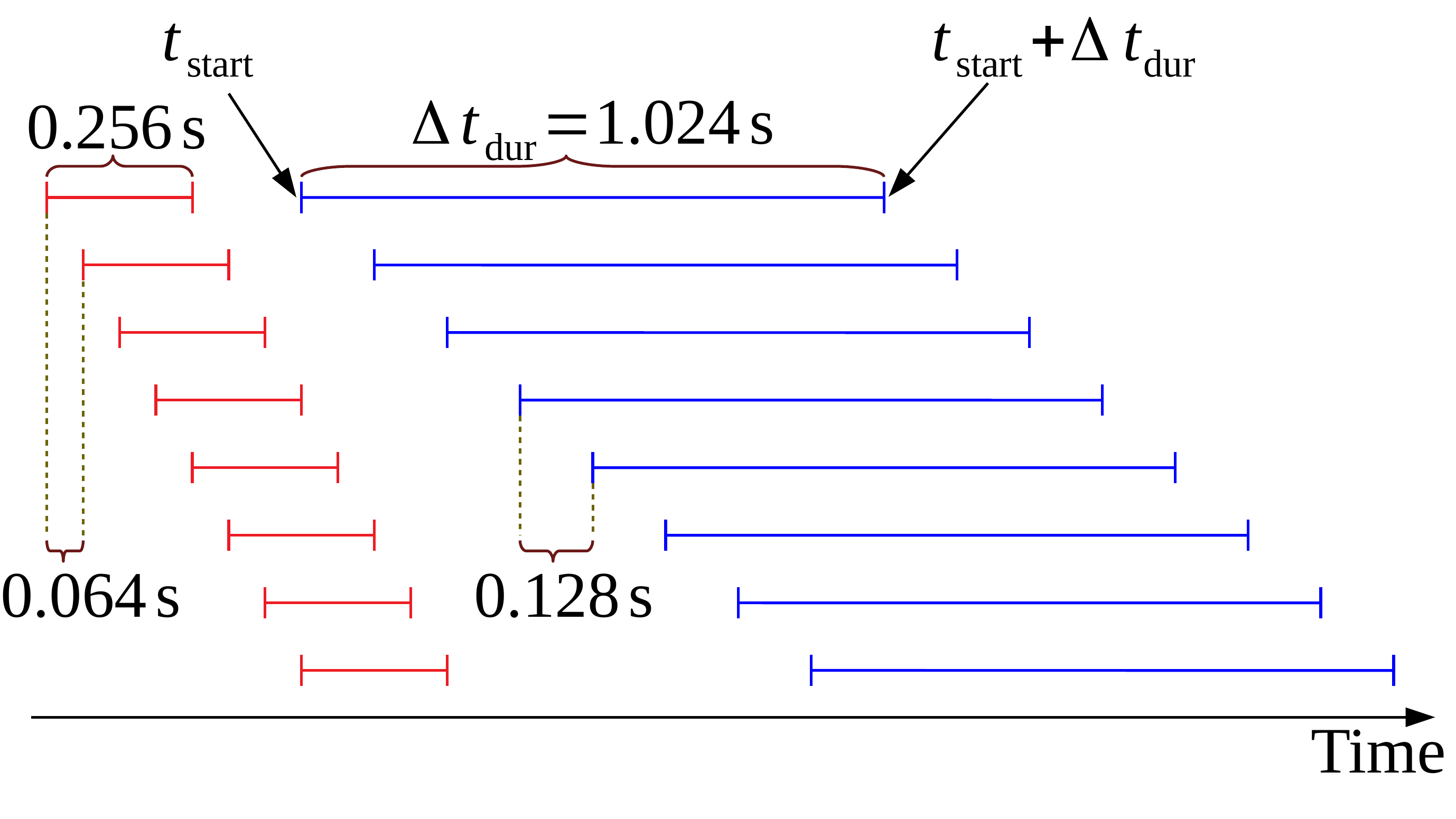}
    \caption{Top panel: time dependence of the orbital phase and positions of $I_k$ intervals in the case $N_{\rm bins} = 3$; the binary has $(m_1, m_2) = (10\,M_\odot, 1.4\,M_\odot)$ and the origin of the x axis coincides with the merger time. Bottom panel: positions of the generic targeted search time windows $[t_{\rm start}, t_{\rm start} + \Delta t_{\rm dur}]$.}
    \label{fig:Ik_and_windows}
\end{figure}

We propose a statistical framework which adapts the search methods presented in \cite{Goldstein:2016zfh, Blackburn:2014rqa}. Hereafter, we refer to the original search as the \emph{generic targeted search}, since it aims at detecting a generic transient excess of high-energy photons above the detector background associated with a particular target time, regardless of its temporal morphology. We refer to our modified method as the \emph{chirp targeted search} instead, as it aims at detecting excesses of photons that exhibit repetitions locked in phase with the orbital evolution of a compact binary that is about to merge. The main difference with respect to the generic targeted search is that we apply its statistical formalism in the orbital phase space, instead of the time space. The transformation from time space to orbital phase space is provided, approximately, by the coalescence time $t_c$ and the component masses $m_1$ and $m_2$ inferred from the GW signal. For a fixed number of bins $N_{\rm bins}$, the orbital phase is split into $N_{\rm bins}$ equal intervals, i.e. $I_0 = \left[0, \frac{2\pi}{N_{\rm bins}}\right]$, $I_1 = \left[\frac{2\pi}{N_{\rm bins}}, 2\frac{2\pi}{N_{\rm bins}}\right]$, \ldots, $I_{N_{\rm bins}-1} = \left[(N_{\rm bins} - 1)\frac{2\pi}{N_{\rm bins}}, 2\pi\right]$.
The top panel of Figure~\ref{fig:Ik_and_windows} illustrates the conversion from the time space to the orbital phase space.
Once this transformation is defined, we can rebin the photons registered by GBM (the TTE data) into the orbital-phase intervals: using each photon's arrival time $t$, we identify its orbital phase interval $I_k$ such that $\Phi_{\rm orbit}(t) \in I_k$.

Next, we need an estimate of the background photon rate in each phase interval, i.e.~the rate of photons registered by GBM in the absence of a modulated transient.
To this end, we first estimate the background rate over time using the unbinned Poisson maximum likelihood technique introduced in \cite{Goldstein:2016zfh}: at a given time $t_0$, the background photon rate $\lambda_{\rm max}(t_0)$ is defined as the ratio between the number of photons $N_{\rm photons}$ contained in a large enough time window of duration $T$ (in our case $T=100\,\text{s}$) and the width of the window, i.e. $\lambda_{\rm max}(t_0) = \frac{N_{\rm photons}}{T}$.
Assuming the background can be described as a stationary Poisson process over the search interval, the uncertainty in its rate can be written as $\sigma^2_{\lambda_{\rm max}}(t_0)= \lambda_{\rm max}(t_0) / T$.
The time window is slid over the time range of interest; thus a background photon rate (assigned with standard deviation) is calculated for an array of times, and finally the photon rates (and their uncertainties) are interpolated over time. A chi-squared statistic $\chi^2$ is computed in order to evaluate the quality of the fit. If the fit is poor, i.e.~$\chi^2$ is too large, for one of the GBM energy channels, that channel is excluded from the search. The background rate is then transformed to the orbital phase space to predict the rate in each phase interval.

Once the foreground photon histogram (background photon fitting) is calculated (estimated) for the $[0, 2\pi]$ orbital phase range, we aim to search for a subset of adjacent intervals $I_k$ presenting an excess of photons. Such a behaviour is equivalent to saying that the binary, during an orbit, emits the majority of the radiation in a specific orbital phase window. This feature is characteristic to the simplified lightcurves. 

A quantity combining information about both source and noise is the likelihood ratio, defined as
\begin{equation}
    \Lambda(d) = \frac{P(d|H_1)}{P(d|H_0)},
\end{equation}
where $d$, $H_1$ and $H_0$ are the observed data, the signal presence hypothesis, and the hypothesis of noise alone. As in~\cite{Blackburn:2014rqa}, the assumption of uncorrelated Gaussian noise allows us to write the preceding probabilities in the following way:
\begin{eqnarray}
    P(d|H_1, s) & = & \prod_i \frac{1}{\sqrt{2\pi}\sigma_{d_i}}\exp{\left(-\frac{(\tilde{d}_i - r_i s)^2}{2\sigma_{d_i}^2}\right)}, 
    \label{eqn:prob_foreground}\\
    P(d| H_0) & = & \prod_{i}\frac{1}{\sqrt{2\pi}\sigma_{n_i}}\exp{\left(-\frac{\tilde{d}_i^2}{2\sigma_{n_i}^2}\right)}.
    \label{eqn:prob_background}
\end{eqnarray}
In those previous formulas, $\tilde{d}_i = d_i - <n_i>$, where $d_i$ and $<n_i>$ are the foreground and the estimated background photons. $\sigma^2_{n_i}$ and $\sigma^2_{d_i}$ represent the variances of the background and the expected data (background plus signal). The variances appearing in Equations~\ref{eqn:prob_foreground} and~\ref{eqn:prob_background} are computed in the orbital phase space, and are obtained by the summation of time space variances. Lastly the time space standard deviations are calculated as in~\cite{Blackburn:2014rqa}.
$r_i$ stands for the detector-energy response, which depends on both the EM source sky location and spectrum. Finally $s$ is the amplitude (measured by the Earth) of the signal. Moreover, for all these quantities, the index $i$ designates a pair (detector, energy channel). Given that $\sigma_{d_i}$, $\sigma_{n_i}$ and $\tilde{d}_i$ are measured quantities, while $r_i$ is calculated for a sample grid, accounting for all possible locations, as explained in~\cite{Kocevski:2018suj}, then the amplitude parameter $s$ is the only variable over which the marginalization needs to be done. Thus, the expression of the likelihood ratio becomes $\Lambda(d) = \int{\frac{P(d|H_1, s)}{P(d|H_0)}P(s)ds}$, where $P(s)$ is the amplitude signal prior. Maximizing the likelihood ratio is the same as maximizing the logarithm of it, $\mathcal{L}(d) = \ln{\Lambda(d)}$. As explained in~\cite{Blackburn:2013ina}, $\ln{\Lambda(d|s)}$ is almost a Gaussian function with variance $\sigma^2_{\ln{\Lambda(d|s)}} = \frac{1}{\sum_{i}r_i^2/\sigma^2_{d_i}}$. Therefore the maximum of $\ln{\Lambda(d|s)}$ is reached for $s_{\rm best}$, obtained by means of the iterative Newton's method. The ${(k+1)}^{\rm th}$ step in the Newton's method consists in refining the $k^{\rm th}$ estimate by using the analytic second derivative, i.e. $s_{k+1} \approx s_k - \frac{\partial \mathcal{L} / \partial s}{\partial^2\mathcal{L}/\partial s^2}$, while the initial guess is $s_0 = \frac{\sum_{i}r_i \tilde{d}_i / \sigma^2_{d_i}}{\sum_{i}r^2_i/\sigma^2_{d_i}}$.
We assume the same well-behaved prior like in~\cite{Kocevski:2018suj}, i.e.
\begin{equation}
    P(s) = \left[1 - \exp{\left(-\left(\frac{s}{\gamma_{\rm prior}\sigma_{\ln{\Lambda(d|s)}}}\right)^{\beta_{\rm prior}}\right)}\right]s^{-\beta_{\rm prior}}, \\
\end{equation}
where $\gamma_{\rm prior}=2.5$ and $\beta_{\rm prior}=1$. The value of $\gamma_{\rm prior}$ ensures a prior almost constant over a range of $\sigma_{\mathcal{L}}$, while the value of $\beta_{\rm prior}$ translates in a luminosity distribution independent of distance. The log-likelihood ratio becomes
\begin{widetext}
\begin{equation}  
\mathcal{L}(d) = \ln{\sigma_{\ln{\Lambda(d|s)}}} + \ln{\left[1 + \text{erf}\left(\frac{s_{\rm best}}{\sqrt{2}\sigma_{\ln{\Lambda(d|s)}}}\right)\right]} 
+ \ln{\Lambda(d|s_{best})} 
+
\left\{\begin{array}{ll}
        \ln{\left[1 - \exp{\left(-\frac{s_{\rm best}}{\gamma_{\rm prior} \sigma_{\ln{\Lambda(d|s)}}}\right)}\right]} - \beta_{\rm prior}\ln{s_{\rm best}}  &  \text{if } s_{\rm best}\geq 0 \\
        \\
        -\beta_{\rm prior}\ln{\left(\gamma_{\rm prior} \sigma_{\ln{\Lambda(d|s)}}\right)} &
        \text{if }  s_{\rm best}\leq 0\\
\end{array}.\right.
\label{eqn:log-likelihood}
\end{equation}
\end{widetext}
In Equation~\ref{eqn:log-likelihood}, $\text{erf}(x) = \frac{2}{\sqrt{\pi}}\int_{0}^{x}\exp{\left(-t^2\right)}dt$ is the error function. And finally, as explained in~\cite{Blackburn:2014rqa}, we calibrate the log-likelihood ratio (hereafter LLR) by subtracting the quantity $\mathcal{L}_{\rm ref} = \beta_{\rm prior}\ln{\gamma_{\rm prior}} + (1 - \beta_{\rm prior})\ln{\sigma_{\rm ref}}$. Here we do not care about the value of $\sigma_{\rm ref}$ because $\beta_{\rm prior}=1$, and so the term $(1-\beta_{\rm prior})\ln{\sigma_{\rm ref}}$ cancels. 

Various kinds of transients commonly appear in Fermi-GBM data and produce large LLR values, despite being certainly unrelated to GW events.
One such class is represented by high-energy cosmic rays hitting the NaI detectors, being responsible for long-lived phosphorescent light emission.
By means of the technique introduced in~\cite{Blackburn:2014rqa}, we remove most undesirable high-LLR triggers from this class.
A second class is represented by sharp photon-rate changes due to Fermi approaching the South Atlantic Anomaly (SAA).
These are discarded instead as explained in \cite{Goldstein:2016zfh}.

Both the generic targeted search and our chirp targeted search take as input a GPS time $t_c$, which is the compact binary merger time, measured at Fermi.
In this work, the generic targeted search is performed over the following exact timescales: $0.064\,\text{s}$, $0.128\,\text{s}$, $0.256\,\text{s}$, $0.512\,\text{s}$, $1.024\,\text{s}$, $2.048\,\text{s}$, $4.096\,\text{s}$ and $8.192\,\text{s}$. 
It uses a time displacement of $64\,\text{ms}$ for the four shortest timescales and a time displacement of factor 8 for the four longest timescales (e.g., the $8.192\,\text{s}$ search windows are separated by $1.024\,\text{s}$).
Thus, in total there are $2270$ such windows.
The positions of some search windows on the time axis are shown in the bottom panel of Figure~\ref{fig:Ik_and_windows}.
For the chirp targeted search, we shorten/extend the same search windows by a small amount in such a way that each window contains an integer number of orbits.
This procedure is important in order to avoid artificially unequal number of photons in different $I_k$ intervals, which would produce artificially large LLR values.
Additionally, in the case of the chirp targeted search, for a fixed $(t_{\rm start}, \Delta t_{\rm dur})$, searches are done over 47 subsets of adjacent intervals $I_k$.
More precisely, we consider subsets of any length in $\left\{1, 2,..., N_{\rm bins}\right\}$ and we use a phase factor of 2 (e.g.~subsets of length 2, 3, and 6 are separated by 1, 2 and 3 intervals $I_k$).
Finally the most significant trigger, i.e. with the highest LLR, is reported.

\subsection{Null distribution of LLR}
\label{sec:llr_noise}
\begin{figure*}[htb]
    \begin{center}
    \includegraphics[width=\columnwidth]{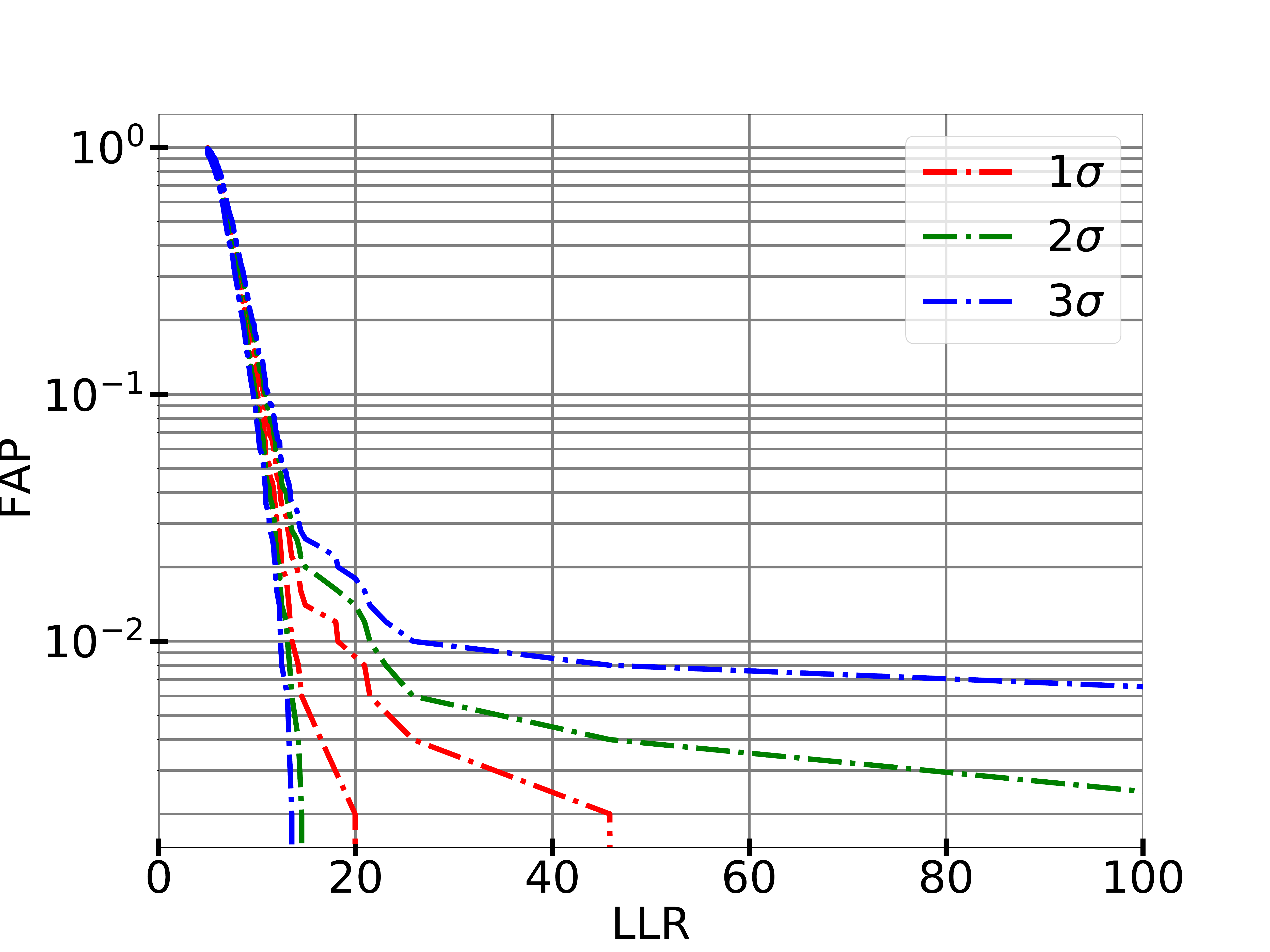}
    \includegraphics[width=\columnwidth]{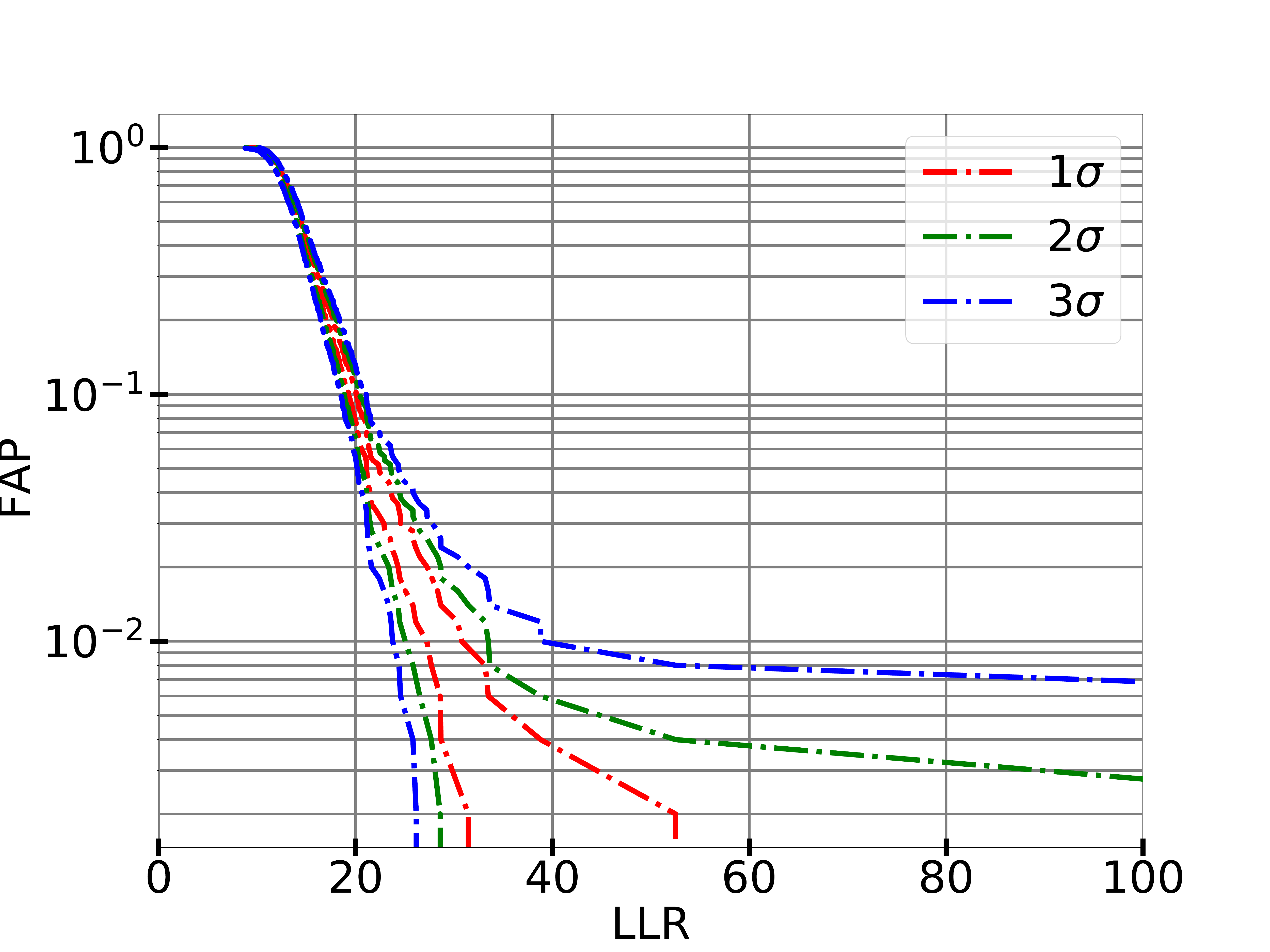}
    \end{center}
    \begin{center}
    \includegraphics[width=\columnwidth]{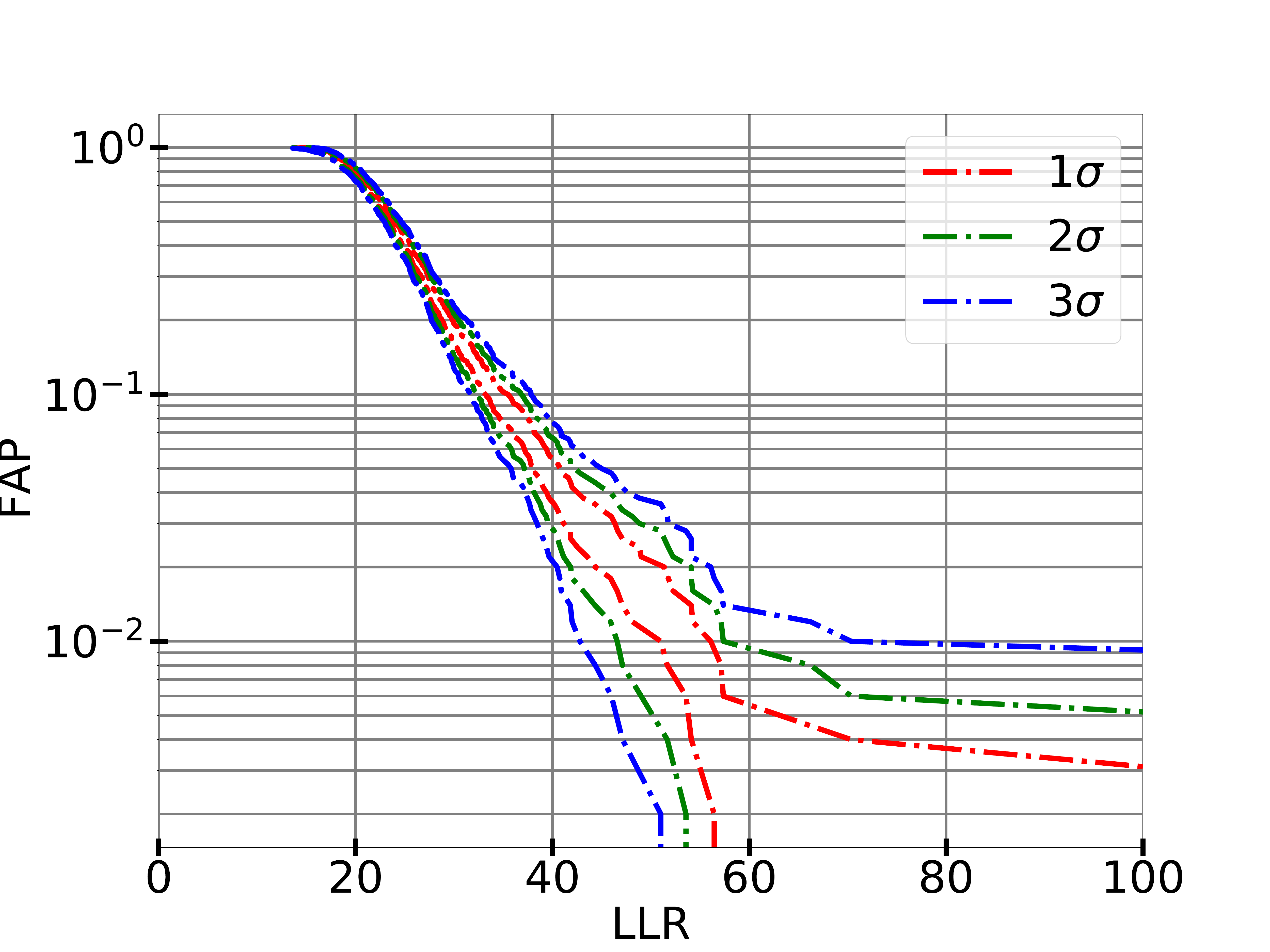}
    \includegraphics[width=\columnwidth]{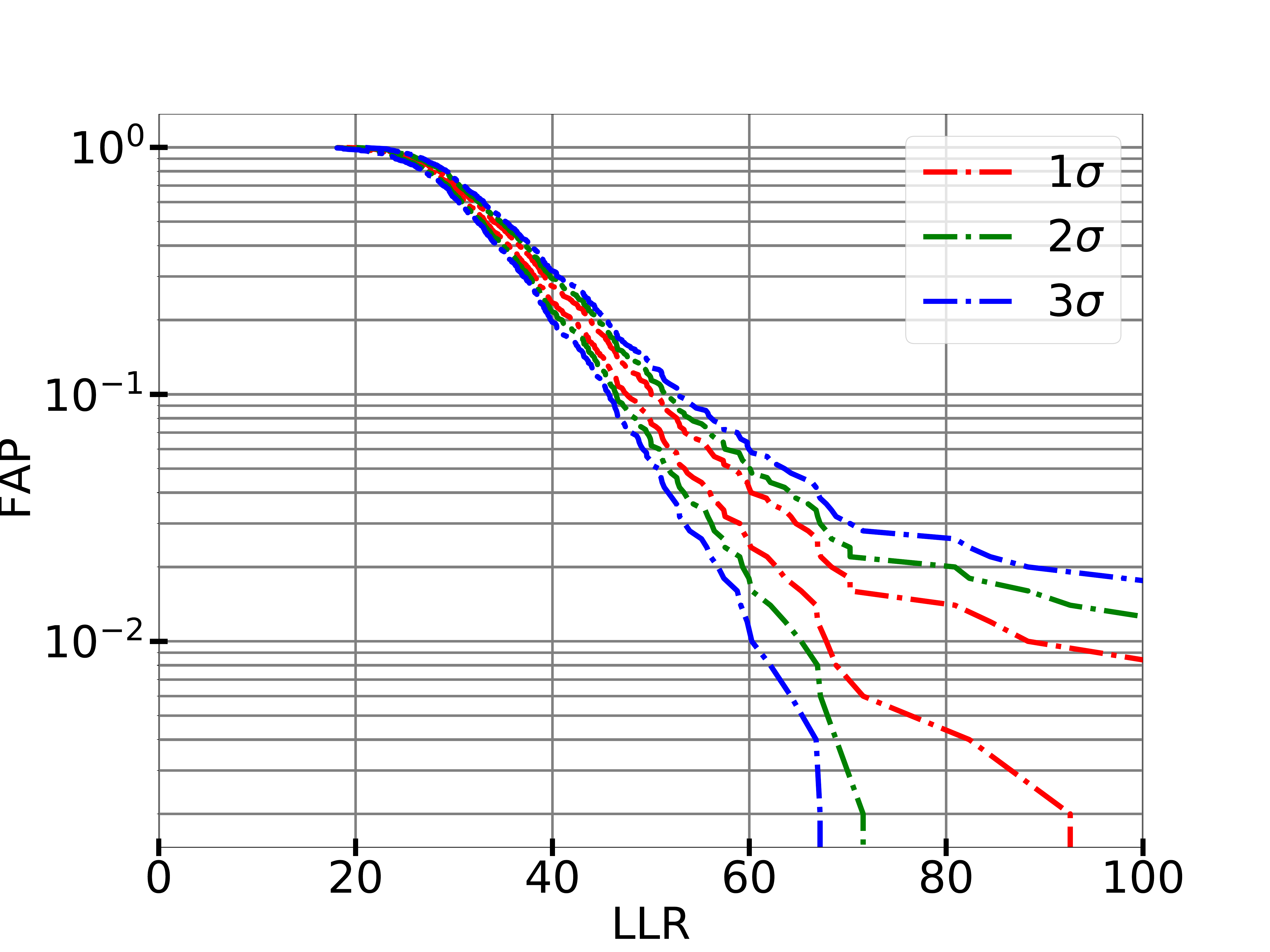}
    \end{center}
    \caption{FAP versus LLR of triggers by trying the two searches on random times. Plotted are the background distributions assigned with $\pm 1\sigma, \pm 2\sigma, \pm 3\sigma$ uncertainties. The top left panel corresponds to the distribution of the generic targeted search output triggers, while the top right, bottom left and bottom right panels represent the chirp targeted search distributions with settings $N_{\rm bins} = 5, 10$ and $15$. For all panels of the chirp targeted search, the setting $(m_1, m_2) = (1.6\,M_\odot, 1.4\,M_\odot)$ is used.}
    \label{fig:background_llr}
\end{figure*}

It is worth mentioning that a statistically significant trigger is not necessarily a GW-related signal, and sometimes not even the effect of an EM radiation intercepted by the NaI and/or BGO detectors. Despite the filtering strategy discussed in Section~\ref{sec:statistical_method}, large LLR spurious signals survive. For this reason, an empirical measure of the false alarm probability (FAP) distribution is extremely useful. In this paper, for a given log-likelihood ratio ${\rm LLR}_0$, the value of ${\rm FAP}({\rm LLR}_0)$ represents the fraction of noise events in which at least one set member has a statistical significance higher than ${\rm LLR}_0$. 
In Figure~\ref{fig:background_llr}, we show the FAP distribution of LLR, obtained by running the search on 1000 random times, spread over the period going from April 1, 2019 to April 1, 2020.
The random times have been chosen such that they are at least $30\,\text{s}$ away from the SAA entrance/exit. This choice is motivated by the exclusion of those situations where the $30\,\text{s}$ time windows would otherwise analyze non-science times.
Based on Figure~\ref{fig:background_llr}, at least two important remarks should be made: (i) even though similar, the distributions corresponding to the chirp targeted search have higher LLRs with respect to the generic-targeted-search distribution; (ii) the noise output triggers have higher LLR as the number of bins $N_{\rm bins}$ increases. 
The remark (i) is an expected behavior, because the generic targeted search is included in the chirp targeted search. In fact, the evaluation of the statistical significance of all $I_k$ intervals together is equivalent to performing the generic targeted search. Regarding (ii), there are at least two reasons favoring this behavior: (a) for two bins numbers $N_{\rm bins, 1}$ and $N_{\rm bins, 2}$, with $N_{\rm bins, 2} > N_{\rm bins, 1}$, such that $N_{\rm bins, 2}$ is a multiple of $N_{\rm bins, 1}$, the chirp targeted search with setting $N_{\rm bins, 1}$ is included in the chirp targeted search with setting $N_{\rm bins, 2}$; (b) the higher the number of bins $N_{\rm bins}$, the less true is the approximation of Gaussian background noise in the high energy detectors, on the scale of a $I_k$ interval.

\section{Search sensitivity}
\label{sec:sensitivity}
In this section, we test the sensitivity of our chirp targeted search and we compare it to the sensitivity of the generic targeted search.
In the case of modulated $\gamma$-ray signals, in phase with GWs, we expect better performance for the chirp targeted search.

\subsection{Properties of the simulated signals}

In this subsection, we describe the simplified lightcurve injections in the GBM data. We consider signals with durations $\Delta t_{\rm dur}$ log-uniformly distributed in $[0.064\,\text{s}, 8.192\,\text{s}]$. For a fixed $\Delta t_{\rm dur}$, the beginning of the radiation, $t_{\rm start}$, is sampled uniformly in $[t_c - 30\,\text{s}, t_c -\Delta t_{\rm dur}]$. 
The injections are uniformly distributed in the sky with $\theta_{\rm ra} \in [-90^\circ, 90^\circ]$ and $\theta_{\rm dec} \in [-180^\circ, 180^\circ]$, while the spectral index $\kappa$ is uniform in $[0, 2]$. For all the injections, we fix $\theta_{\rm width}  = 10\,^\circ$, while $\theta_{\rm peak}$ is random in $[0\,^\circ, 360\,^\circ]$. The lightcurve amplitude factor $f_{\rm amp}$ is uniformly distributed in $[20\sqrt{\frac{0.064\,\text{ms}}{\Delta t_{\rm dur}}}, 50\sqrt{\frac{0.064\,\text{ms}}{\Delta t_{\rm dur}}}]$. This choice was found empirically in order to respect the following requirements: the lower and upper limits impose for the majority of the injected signals to have statistical significance right above the background LLR distribution; the dependence on $\Delta t_{\rm dur}$ causes signals with different durations to have similar LLRs.
Finally, all the modulated signals considered in this study are injected at random times spread over one year period starting at April 1, 2019 in such a way that the merger time is always at least $\pm 30\,\text{s}$ away from the closest SAA episode. 

\subsection{Comparison with the generic targeted search}

\begin{figure*}[htb]
    \begin{center}
    \includegraphics[width=\columnwidth]{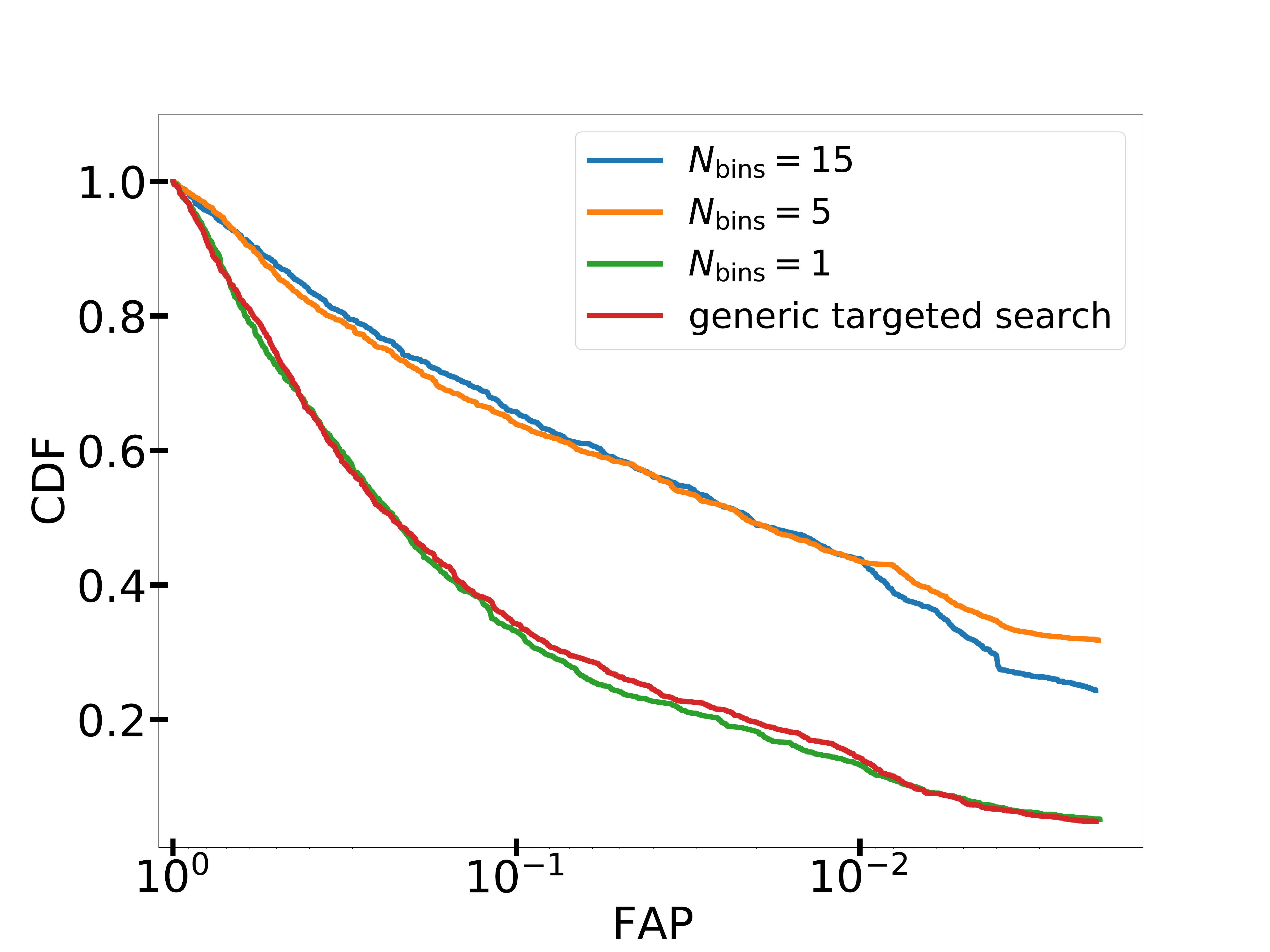}
    \includegraphics[width=\columnwidth]{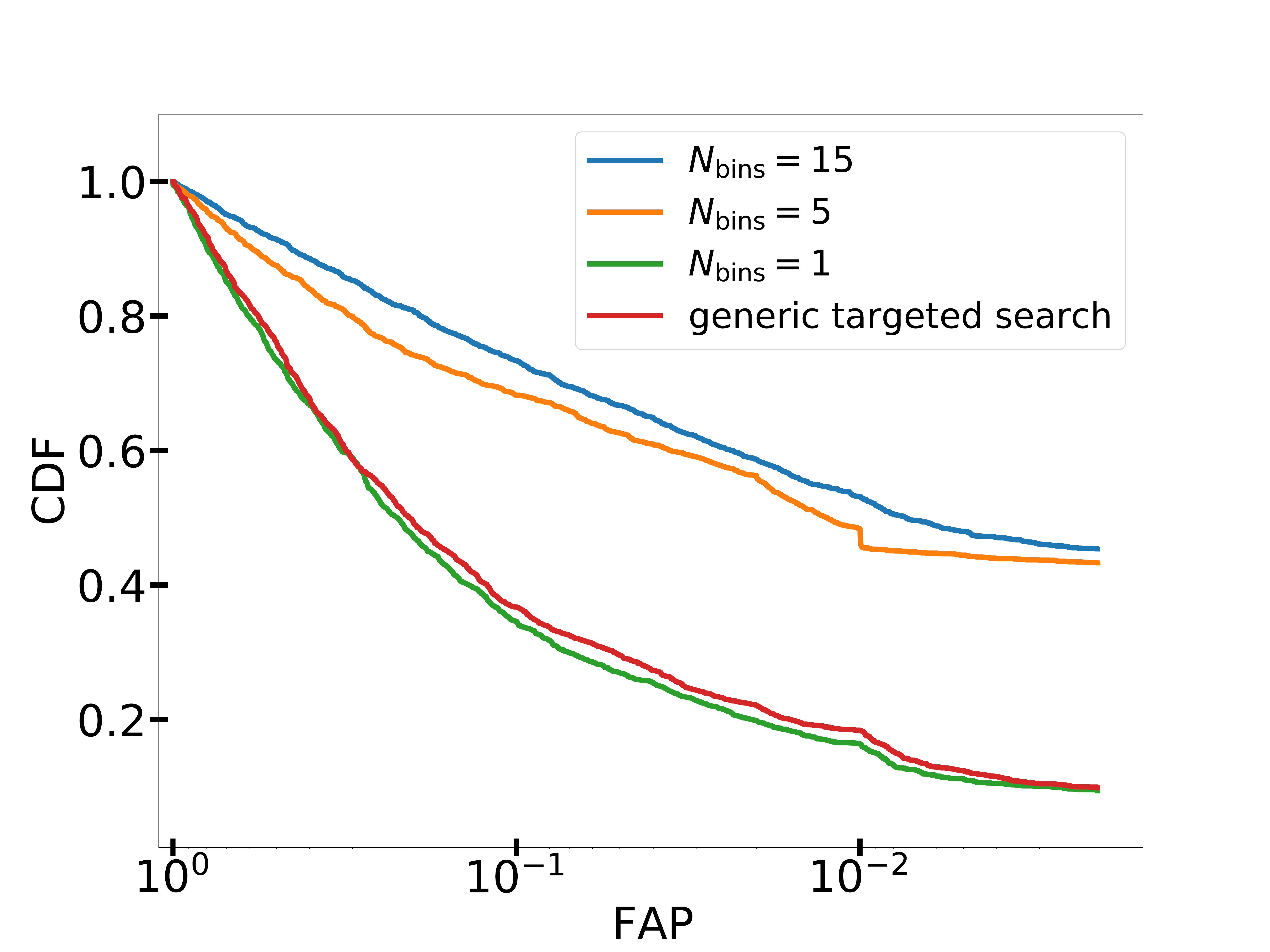}
    \end{center}
    \caption{Fraction of detected injections versus FAP for the simplified lightcurve model. The left (right) panel corresponds to injections with $(m_1, m_2) = (1.6\,M_\odot, 1.4\,M_\odot)$ (respectively $(m_1, m_2) = (10\,M_\odot, 1.4\,M_\odot)$). The injected signals are recovered with the generic targeted search and the chirp targeted search having the correct $(m_1, m_2)$ setting, while $N_{\rm bins}$ is varied.}
    \label{fig:far_foreground}
\end{figure*}

In this subsection, we simulate signals with $(m_1, m_2)$ equal to $(10\,M_\odot, 1.4\,M_\odot)$ and $(1.6\,M_\odot, 1.4\,M_\odot)$, and recover them with the chirp targeted search, where  $(m_1, m_2)$ is fixed at the same value. We also apply the generic targeted search to these injection sets. For each injection, the most significant trigger is selected and a FAP is derived according to the results summarized in Figure~\ref{fig:background_llr}. The cumulative distribution functions (CDF) of the FAP for these injections sets are illustrated in Figure~\ref{fig:far_foreground}.
The first remark one should draw from Figure~\ref{fig:far_foreground} is that the chirp targeted search is more sensitive than the generic targeted search in the case of simplified lightcurves, as long as $N_{\rm bins} > 1$.
The case of the chirp targeted search with setting $N_{\rm bins}=1$ should be equivalent to the case of the generic targeted search.
This relation is verified here, the small discrepancy between the two distributions being due to the different technical implementations.

The $N_{\rm bins}$ dependence of the chirp targeted search sensitivity warrants some discussion. Firstly, Figure~\ref{fig:background_llr} suggests that the statistical significance of the noise triggers increases with $N_{\rm bins}$, which has as effect the degradation of the sensitivity with the augmentation of $N_{\rm bins}$. Secondly, the performance of the pipeline is expected to depend on the relation between the width of intervals $I_k$ and the orbital phase length of the chirping signals we want to detect. Following this idea, the sensitivity should increase with $N_{\rm bins}$, as long as the width of $I_k$ is higher than the orbital phase length of the recovered signal. 
Indeed, for a given EM chirp radiation, if the signal is included in only one interval $I_k$, the smaller the $I_k$ the higher the signal-to-noise ratio, because the higher the percentage of $I_k$ where the foreground photon rate is above the background photon rate. 
Therefore, one should expect a compromise between the two regimes described above: an increase of the sensitivity with the number of $I_k$ intervals at low $N_{\rm bins}$, then a saturation followed by a degradation of the performance at high $N_{\rm bins}$. This behavior is verified in Figure~\ref{fig:far_foreground}. 
For $N_{\rm bins}$ equal to $1$, $5$ and $15$, the width of the interval $I_k$ is $360\,^\circ$, $72\,^\circ$ and $24\,^\circ$. Given that $\theta_{\rm width} = 10\,^\circ$, such a signal could a priori be included in one interval $I_k$, unless it is situated at the border of two adjacent $I_k$ intervals. From Figure~\ref{fig:far_foreground}, one can note an increase of the sensitivity between the cases $N_{\rm bins} =1$ and $N_{\rm bins}=5$. However, the performances of the pipeline seem to be quite similar in the cases $N_{\rm bins} = 5$ and $N_{\rm bins} = 15$.

\subsection{Detectability of signals with different parameters}

\begin{figure*}[htb]
    \begin{center}
    \includegraphics[width=\columnwidth]{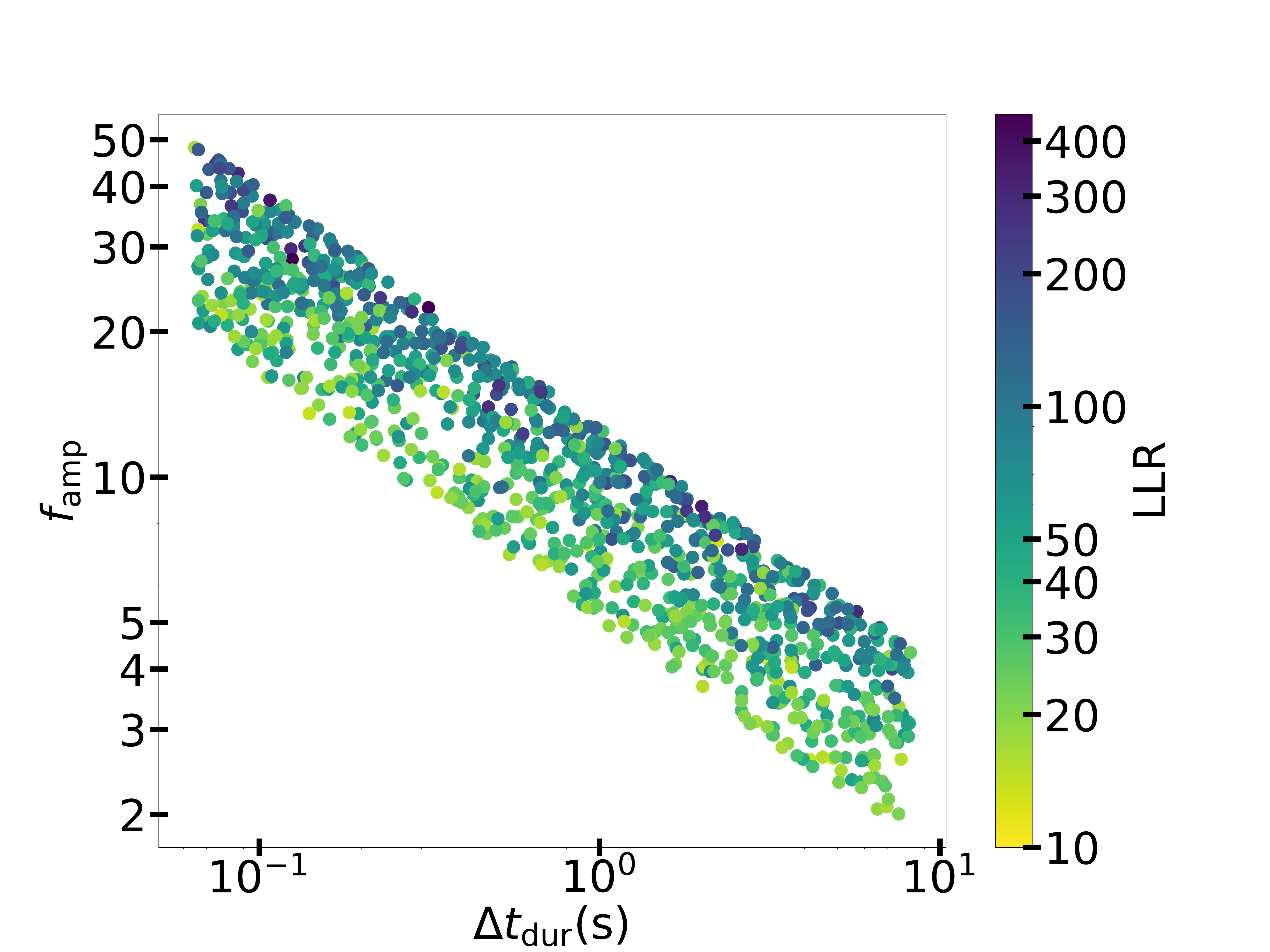}
    \includegraphics[width=\columnwidth]{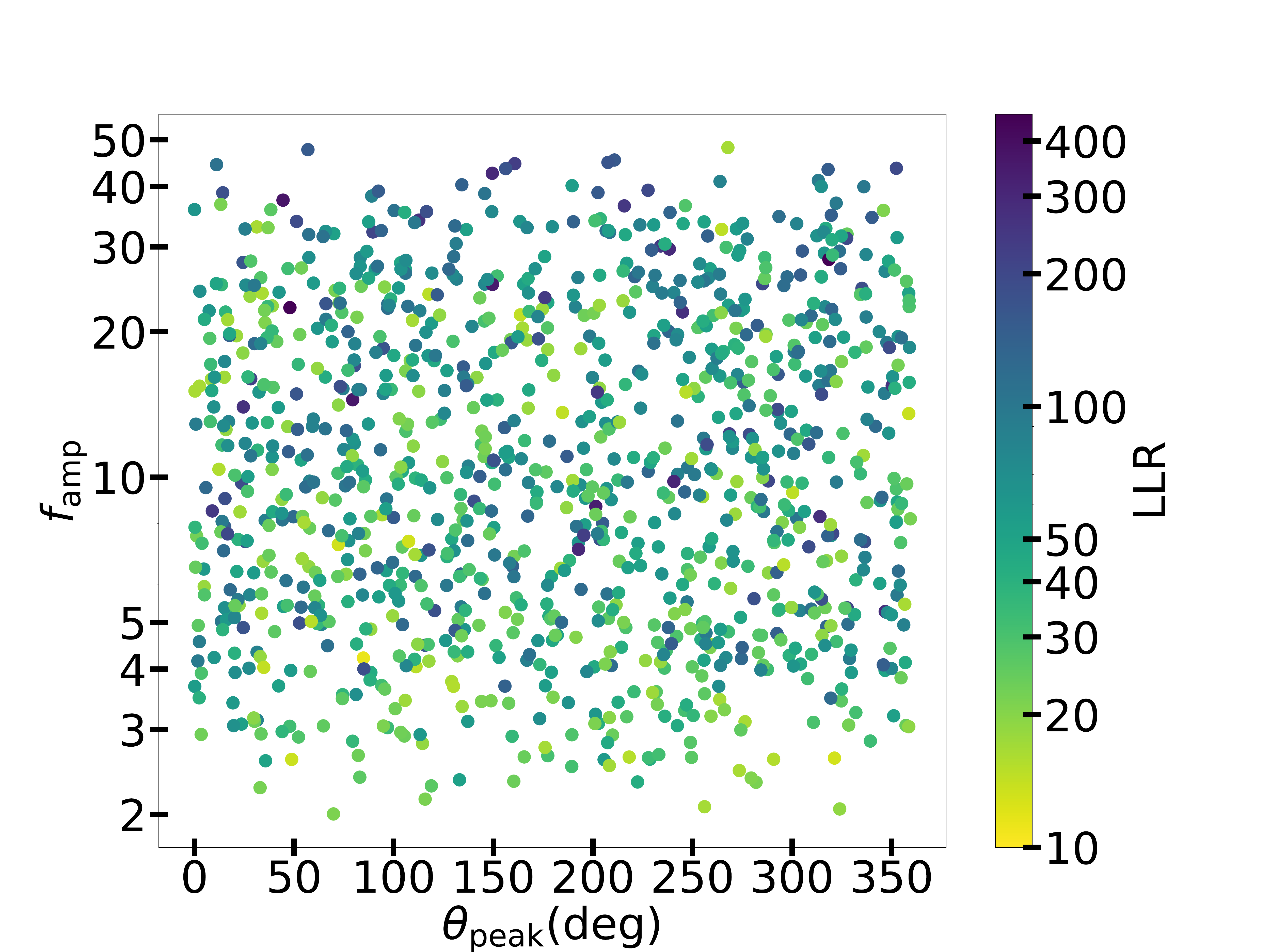}
    \end{center}
    \begin{center}
    \includegraphics[width=\columnwidth]{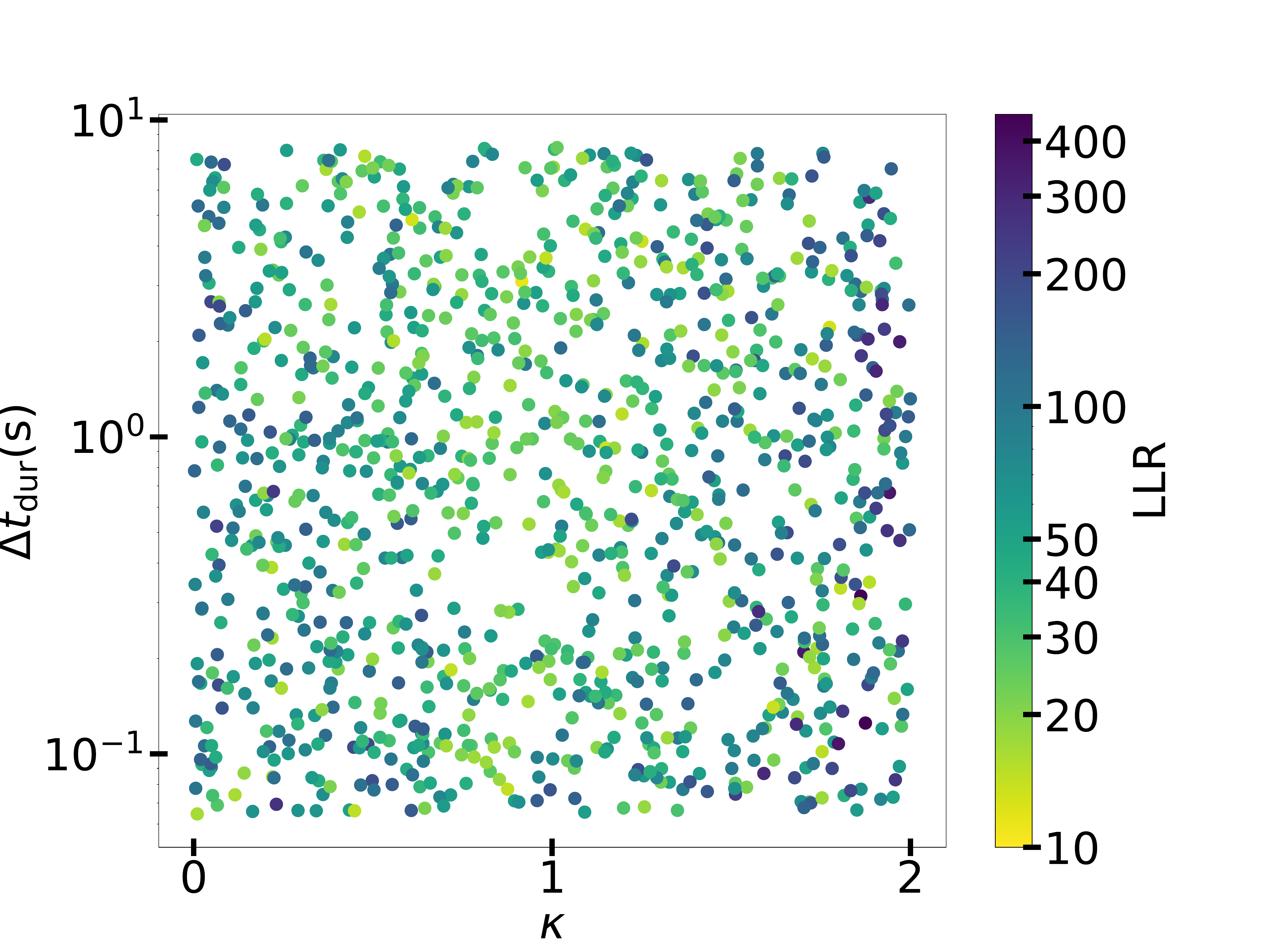}
    \includegraphics[width=\columnwidth]{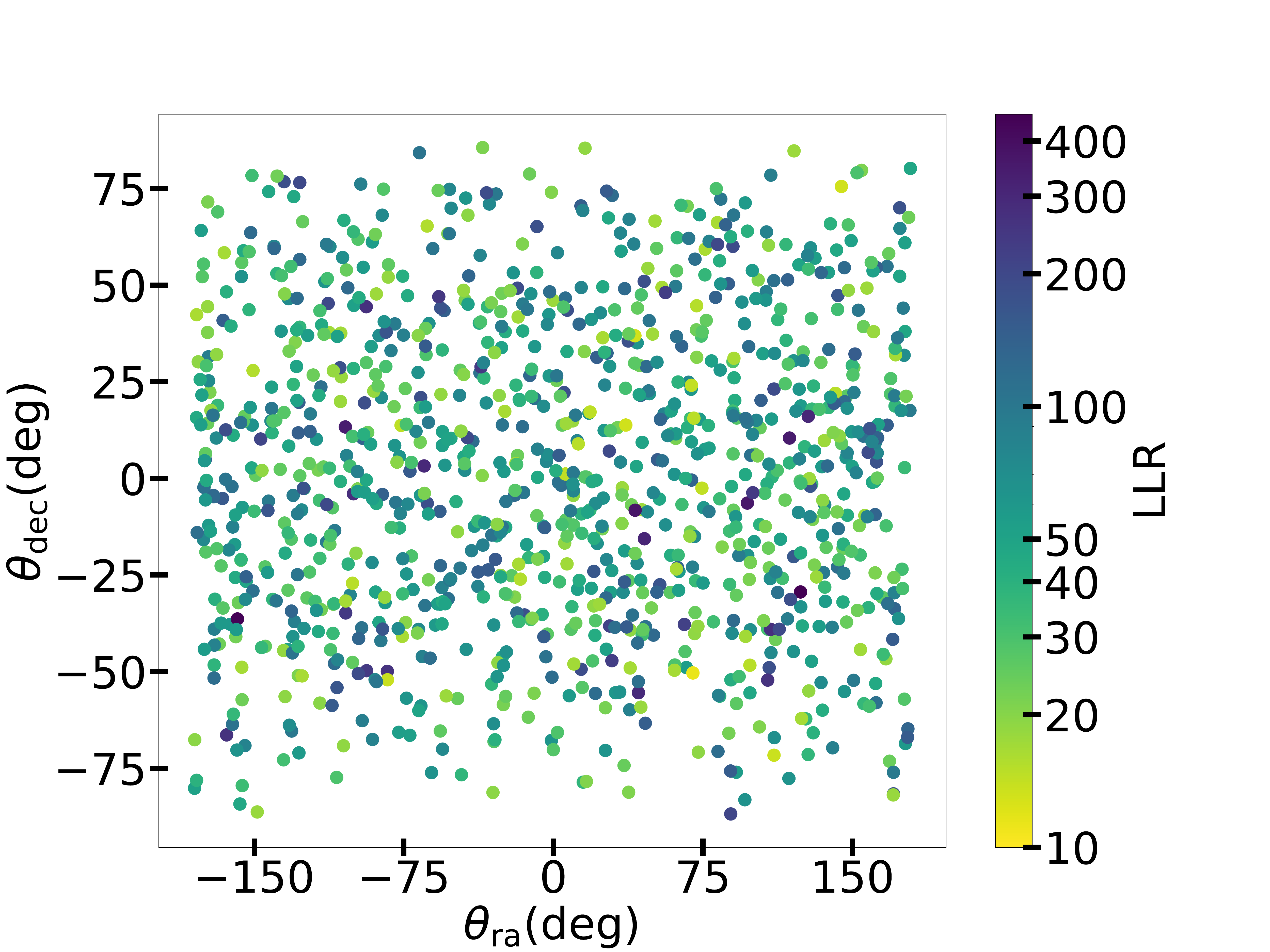}
    \end{center}
    \caption{Variation of the LLR statistic produced by the search with the parameters of the simulated simplified lightcurves. All the panels correspond to the injection set with $(m_1, m_2) = (10\,M_\odot, 1.4\,M_\odot)$, while the merger times are uniformly distributed over the one year period starting on April 1, 2019. The chirp targeted search is used with setting $N_{\rm bins} = 10$.}
    \label{fig:parameters_exotic}
\end{figure*}

Figure~\ref{fig:parameters_exotic} shows the statistical significance dependence of the chirp targeted search with the parameters of the simplified-lightcurve injections. 
The top left panel of Figure~\ref{fig:parameters_exotic} proves that our choice for the $\Delta t_{\rm dur}$ dependence of the amplitude factor $f_{\rm amp}$ puts on an equal footing the signals with different durations. One should note that an injection with $f_{\rm amp}=1$ corresponds to an  energy flux of $1\,\text{erg/s/}\text {cm}^2$ in the $50-300\,\text{keV}$ band. Thus, the same panel indicates that a signal spread over around $1\,\text{s}$ and emitting during $10\,^\circ$ orbital phase window, with a flux reaching  $10\,\text{erg/s/}\text {cm}^2$ in the Fermi-GBM band, might be recovered by our pipeline with an important statistical significance ($\text{LLR} > 100$).
The top right panel shows that the pipeline sensitivity is independent of the signal position in the orbital phase space. From the bottom left panel, one can note better performance of the chirp targeted search when the signal has a spectrum either close to $\kappa=0$ (\textit{hard}) or close to $\kappa=2$ (\textit{soft}). This might be explained by the fact that although for the injections we have considered $\kappa$ anywhere in the interval $[0, 2]$, the search recovery is realized with only the three spectrum templates corresponding to $\kappa \in \{0, 1, 2\}$. Therefore, an injection with $\kappa$ not near to 0 or 2, represents an heterogeneous weighted addition of quite different spectral templates, and which is more difficult to be recovered with only one template in between $\{$\textit{soft}, \textit{normal}, \textit{hard}$\}$. Finally, the bottom right panel of Figure~\ref{fig:parameters_exotic} illustrates the independence of the sensitivity with the position in the sky of the injected signal.

\subsection{Impact of orbital-phase uncertainty}

The properties of a compact binary merger inferred from the GW data always carry some uncertainty. In particular, there are uncertainties on the chirp mass, the merger time and the sky location. The chirp mass uncertainty will impact our knowledge about the orbital phase evolution. The uncertainties on the geocentric merger time and the sky location together reflect on the uncertainty of the merger time at the position of the Fermi satellite, which in turn affects our knowledge of the positions of the $I_k$ intervals. Concerning the chirp mass detection errors, the O1, O2 and O3 LIGO-Virgo observing runs showed that the uncertainties are below $0.1\,M_\odot$ for BNS or NSBH like objects, and of the order of a few $M_\odot$ for binary black hole mergers. Concerning the geocenter merger time detection, the uncertainty is of the order of $1\,\text{ms}$. However the sky locations of GW mergers are often poorly constrained, especially in the case of single-interferometer detections, which means that the uncertainty of the merger time measured at Fermi is of the order of the photon flight time between Fermi and the center of the Earth, i.e.~$\approx 23$ ms.
Before applying our method to LIGO-Virgo events, we have to evaluate the impact of imprecise GW measurements on the sensitivity of our search.

\begin{figure}[tb]
    \includegraphics[width=\columnwidth]{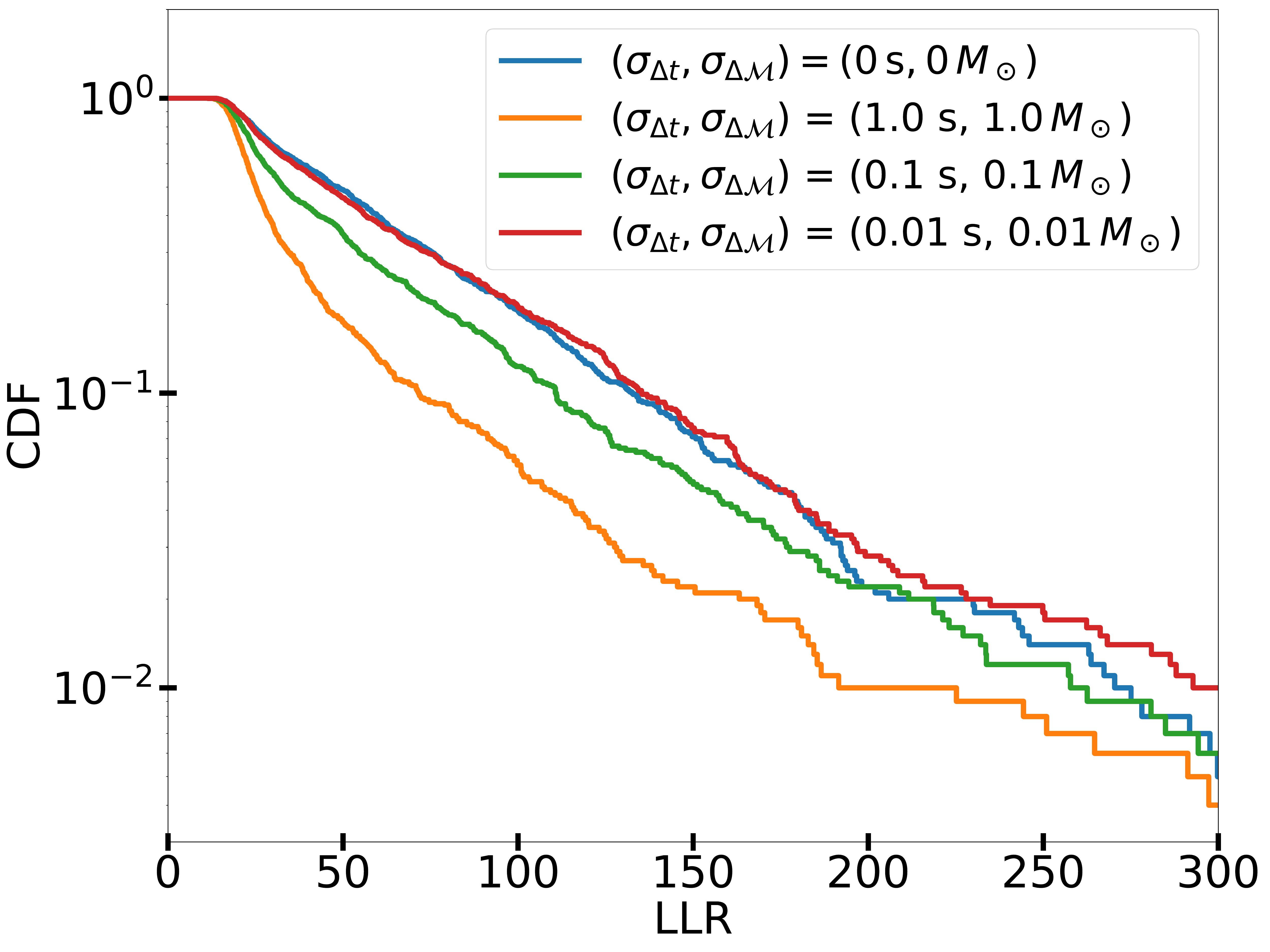}
    \caption{CDF versus LLR for simplified-lightcurve injections assigned with merger time and chirp mass uncertainties $\sigma_{\Delta t}$ and $\sigma_{\Delta\mathcal{M}}$.  The injections are uniformly spread over the period between April 1, 2019 and April 1, 2020, and the setting $N_{\rm bins}=10$ is used.} 
    \label{fig:with_uncertainty}
\end{figure}

The variation of the sensitivity with the uncertainty on the Fermi location merger time and on the chirp mass is illustrated in Figure~\ref{fig:with_uncertainty}.
In order to investigate the impact of the simultaneous inaccuracies of the merger time and chirp mass measurements on the recovery efficiency of the search, we use a set of 1000 injections. Given the total mass $M_{\rm total} = m_1 + m_2$ and the chirp mass $\mathcal{M}$, we consider injections with $(t_c, M_{\rm total}, \mathcal{M}) = (t_c^i + \delta t_c^i, M_{\rm total}^0, \mathcal{M}_0 + \delta \mathcal{M}^i)$, where $M_{\rm total}^0$ and $\mathcal{M}_0$ correspond to a binary formed of a $10\,M_\odot$ black hole and a $1.4\,M_\odot$ neutron star, while $\delta t_c^i$ and  $\delta \mathcal{M}^i$ are uniformly sampled in $[-\sigma_{\Delta t}, \sigma_{\Delta t}]$ and $[-\sigma_{\Delta \mathcal{M}}, \sigma_{\Delta \mathcal{M}}]$.
Then we recover the injected signal by means of the chirp targeted search, with the setting $(t_c^i, M_{\rm total}^0, \mathcal{M}_0)$.
This is equivalent to a GW trigger having a merger time uncertainty at Fermi of $\sigma_{\Delta t}$ and a chirp mass uncertainty equal to $\sigma_{\Delta \mathcal{M}}$.
In the previous expressions the exponent $i$ stands for the $i^{\rm th}$ trigger.
According to the results presented in Figure~\ref{fig:with_uncertainty}, when used with uncertainties $\sigma_{\Delta t} \leq 0.01\,\text{s}$ and $\sigma_{\Delta \mathcal{M}} \leq 0.01\,M_\odot$, the CDF chirp targeted search has a relative error with respect to the case of perfect measurements, i.e.~$(\sigma_{\Delta t}, \sigma_{\Delta \mathcal{M}}) = (0\,\text{s}, 0\,M_\odot)$, lower than $10\%$ over the range $\text{LLR} \in [0, 200]$.
This means that if we have a measurement with such low uncertainties, and if the output of the chirp targeted search, when used with setting $N_{\rm bins} \leq 10$, indicates a trigger with $\text{LLR} \geq 100$, then the EM candidate is promising, because it is above the LLR background range corresponding to $-1\sigma$ lower limit.

\section{Results on LIGO-Virgo detections} 
\label{sec:real_events}

We finally apply the search method described previously to a few LIGO-Virgo detections which could plausibly be associated with a modulated $\gamma$-ray counterpart: the BNS mergers GW170817~\citep{PhysRevLett.119.161101} and GW190425~\citep{Abbott_2020_190425}; GW190814~\citep{Abbott_2020_190814}, whose heavier object is a black hole, while the lighter object is either the heaviest neutron star or the lightest black hole observed to date; and the neutron star-black hole mergers GW200105 and GW200115~\citep{LIGOScientific:2021qlt}.
Binary black hole mergers, although much more plentiful, are not considered as likely sources of detectable $\gamma$-ray counterparts in this work, and are not investigated. Moreover, the search for EM counterparts to such binary systems would impose a large trials factor due to the high rate of these mergers and the large uncertainties on both the chirp mass and the sky localizations~\citep{PhysRevX.11.021053}. Hence, brighter EM counterparts may be required for a detection. 

Observer-frame chirp masses and mission elapsed times (MET) since  2001.0 UTC (decimal), as well as the corresponding uncertainties, for the five GW events, are given in Table~\ref{tab:PE_parameters}.
Concerning the MET of GW170817 (respectively GW190814),  $3\,\text{ms}$ (respectively $20\,\text{ms}$) have been subtracted from the geocenter merger time posterior, in order to account for the angle between the direction to NGC 4993 (respectively the directions representing $90\%$ credible regions of the GW190814 skymap) and the Fermi-Earth center baseline. For the other events, whose sky localizations are poorly constrained, we make the conservative simplifying assumption that the sky location is completely unknown. We then simply increase (decrease) the upper (lower) limits on the merger time by $23$ ms, which is approximately the light travel time between the Earth center and the Fermi satellite.

\begin{table}[htb]
    \centering
    \begin{tabular}{|c|c|c|}
        \hline
        Event & Merger MET (s) & $\mathcal{M}$ ($M_\odot$)\\
        \hline
        GW170817 & $524666469.424_{-0.002}^{+0.001}$ & $1.198$ \\
        GW190425  &  $577873090.009_{-0.031}^{+ 0.056}$ & $1.487^{+0.001}$ \\
        GW190814 & $587509843.970^{+0.003}$ & $6.413^{+0.012}_{-0.015}$ \\
        GW200105 & $599934271.048_{-0.030}^{+0.058}$ & $3.620_{-0.007}^{+0.009}$ \\
        GW200115 & $600754994.755_{-0.055}^{+0.029}$ & $2.582_{-0.005}^{+0.005}$ \\
        \hline
    \end{tabular}
    \caption{The MET, corresponding to the merger time at Fermi satellite, and the observer-frame chirp mass, for the GW detections explored in this work as possible sources of modulated $\gamma$-ray precursors. The values appearing here are the median (50th percentile), the upper limits (90th percentile) and the lower limits (10th percentile). A missing MET (respectively chirp mass) lower/upper limit means that the limit value is away from the median value by less than $1\,\text{ms}$ (respectively $10^{-3}\,M_\odot$).
    The parameter estimates are given in \cite{romero2020bayesian} for GW170817, \cite{LIGOScientific:2018mvr} for GW190425 and GW190814, and \cite{LIGOScientific:2021qlt} for GW200115 and GW200115.}
    \label{tab:PE_parameters}
\end{table}

To increase our chance of picking merger time and chirp mass values close enough to the true ones, we make use of a grid in the following way: if the MET (respectively the chirp mass) upper and lower limits are $\text{MET}_{\rm min}$ and $\text{MET}_{\rm max}$ (respectively $\mathcal{M}_{\rm min}$ and $\mathcal{M}_{\rm max}$), we consider 2D grid points $(\text{MET}_i, \mathcal{M}_j)$, with $\text{MET}_i$ (respectively $\mathcal{M}_j$) ranging from $\text{MET}_{\rm min}$ to $\text{MET}_{\rm max}$ (respectively from $\mathcal{M}_{\rm min}$ to $\mathcal{M}_{\rm max}$), such that $\text{MET}_{i+1} - \text{MET}_i = 0.02\,\text{s}$ and $\mathcal{M}_{j+1} - \mathcal{M}_j = 0.02\,M_\odot$. In this way, if the true values $\text{MET}^0$ and $\mathcal{M}^0$ are indeed in between the upper and the lower limits, there is at least a grid point $(\text{MET}_{i_0}, \mathcal{M}_{j_0})$ such that $|\text{MET}^0 - \text{MET}_{i_0}| \leq 0.01\,\text{s}$ and $|\mathcal{M}^0 - \mathcal{M}_{j_0}| \leq 0.01\,M_\odot$. 
This working method is motivated by the results obtained in the previous section and summarized in Figure~\ref{fig:with_uncertainty}.
We create the grids corresponding to the five GW events and then we run both the generic and chirp targeted searches with the setting defined by these grid points.

The maximum LLR obtained for each run is converted into a FAP in the following way.
For a given GW event with merger time $t_c$, the chirp targeted search (and the generic targeted search) background distributions are obtained by running the search on 1000 random off-source times covering the interval $[t_c - \frac{1}{2}\,\text{month}, t_c + \frac{1}{2}\,\text{month}]$\textbackslash$[t_c - 60\,\text{s}, t_c + 30\,\text{s}]$.
In this way, the estimated noise distribution samples the high-energy EM activity in the few weeks around the time of interest, but we avoid the $30\,\text{s}$ of on-source Fermi-GBM data preceding and following the merger.
This prevents any possible candidate counterpart from contaminating the background.
Of course, when constructing the off-source background of the chirp targeted search, mass parameters consistent with the particular on-source event of interest are used.

\begin{table}
    \centering
    \begin{tabular}{|c|c|c|c|c|}
     \hline
      & \multicolumn{2}{c|}{Chirp search} & \multicolumn{2}{c|}{Targeted search} \\
      \cline{2-5} Event & \multicolumn{1}{c|}{LLR} &  \multicolumn{1}{c|}{${\rm FAP}^{-3\sigma}_{+3\sigma}$} & \multicolumn{1}{c|}{LLR} &  \multicolumn{1}{c|}{${\rm FAP}^{-3\sigma}_{+3\sigma}$} \\
    \hline
    GW170817 & $52.7$ & $0.034^{-0.021}_{+0.029}$ & $7.0$ & $0.554^{-0.064}_{+0.065}$  \\
    GW190425  &  $26.9$ & $0.290^{-0.061}_{+0.062}$ & $7.2$ & $0.500^{-0.069}_{+0.065}$ \\
    GW190814 & $21.3$ & $0.135^{-0.044}_{+0.049}$ & $8.5$ & $0.217^{-0.055}_{+0.057}$ \\
    GW200105 & $25.4$ & $0.127^{-0.042}_{+0.048}$ & $10.7$ & $0.053^{-0.027}_{+0.033}$ \\
    GW200115 & $24.3$ & $0.237^{-0.057}_{+0.059}$ & $8.9$ & $0.168^{-0.047}_{+0.052}$ \\
    \hline
    \end{tabular}
    \caption{Results of the chirp search and targeted search, in terms of highest LLR and FAP with $\pm 3\sigma$ uncertainties, for the GW events explored in this work. The setting $N_{\rm bins} = 10$ has been used for the chirp search. The FAP values indicate that the LLRs are compatible with background fluctuations.}
    \label{tab:LLR}
\end{table}

The maximum LLR and FAP associated with each GW event are reported in Table~\ref{tab:LLR}.
None of the FAP values have upper limit below 0.01, indicating that the LLR values associated with all events are compatible with the off-source background fluctuations. We conclude that there are neither statistically significant excesses of photons, nor significant modulated signals prior to the mergers we considered.

\section{Conclusion} 
\label{sec:conclusion}
In this work we present a method, the chirp targeted search, to detect modulated $\gamma$-ray precursors to compact binary mergers in Fermi-GBM data. The existence of such signals is not confirmed so far. If they exist, there are several physical mechanisms which might be responsible for their emission. This fact makes the lightcurve amplitude dependence difficult to predict. Despite these difficulties, the presented method is very general. It aims to look for an excess of photons in the orbital phase space, while the GW frequency evolution is defined by the first order term in the post-Newtonian expansion.
The sensitivity of the method has been tested on simplified lightcurves, for which the EM emission takes place only during the same orbital phase window. The performance of the chirp targeted search has been compared to that of an existing, more generic targeted search, which aims to detect an excess of photons in the time space.
It has been displayed that the chirp targeted search has higher sensitivity than the generic targeted search, when the signal is modulated by the GW frequency of the binary.
Finally, both pipelines have been used to search for EM precursors associated with confident GW events having a non-negligible probability to contain a neutron star, namely GW170817, GW190425, GW190814, GW200105 and GW200115.
We found no significant candidate precursor signals associated with any of those events. However, given the potential of new physics provided by the presence of precursor signals, the proposed method, here demonstrated, will be applied to future BNS and NSBH observations by LIGO-Virgo-KAGRA, especially since an increased number of events is expected in the coming years.

It is possible that this method could also be important for stellar-mass binary black hole mergers. As an example, these mergers could occur in gas-rich environments, such as active galactic nuclei disks, and could produce EM counterparts with orbital modulations, similarly to what has been proposed for more massive binaries in the LISA band~\citep[e.g.][]{Tang:2018rfm}.
Potential EM counterparts have been reported for both GW150914~\citep{Connaughton:2016umz} and GW190521~\citep{Graham:2020gwr}.
As stated above, however, in order to apply this method to the large number of binary black hole events detected by current ground-based interferometers, it will be necessary to address the statistical and computational challenges.

Although a sensitivity gain has been achieved by means of the chirp targeted search, improvements can be envisaged in the future.
The actual method is trying to recover signals with spectra described by the related Band function.
One can argue that this is not the most optimal choice.
The Band functions are very appropriate to recover $\gamma$-ray burst prompt emission, at the origin of which most likely the synchrotron radiation and the inverse Compton scattering are at play in producing photons.
In the case of $\gamma$-ray precursors to compact binary mergers, one can imagine other physical emission mechanisms, like thermal emission.
A higher sensitivity might then be obtained by including additional spectral templates in the search.
Additional filtering strategies may also turn out to be effective. 
A cleaning of the search output is synonymous with decreasing the false alarm probability, and as a consequence, the sensitivity of the search gets higher.

\emph{Acknowledgements.} 
TDC was supported by an appointment to the NASA Postdoctoral Program at the
Goddard Space Flight Center, administered by Universities Space Research
Association under contract with NASA, during part of this work.
M.~W.~Coughlin acknowledges support from the National Science Foundation with grant numbers PHY-2010970 and OAC-2117997.

\appendix

\section{EM realistic lightcurve}
\label{sec:realistic_lightcurve}
We present here the analytical formulae used for the derivation of the lightcurve highlighted in orange, in Figure~\ref{fig:waveform_parameters}. The system is assumed to be a NSBH binary and the flaring is due to the uniform and isotropic emission of the neutron star surface. 
The amplitude of the luminosity is modulated by relativistic beaming, gravitational lensing and orbital separation shrinking. 

Concerning the relativistic Doppler beaming, the same magnification factor as in~\cite{dubus2010relativistic} is used. This factor depends on the neutron star velocity $v_{\rm NS}^{\rm DUB}$, the angle $\zeta_1^{\rm DUB}$ between $v_{\rm NS}^{\rm DUB}$ and the line of sight, and the thermal light spectral index $\alpha^{\rm DUB}$. The expression of the amplification factor is

\begin{equation}
    \left[ 
    \frac{1 - \frac{v_{\rm NS}}{c}\cos\left( {\zeta_1^{\rm DUB}} \right)}{\sqrt{1 - \left(\frac{v_{\rm NS}^{\rm DUB}}{c}\right)^2}}
    \right]^{\alpha^{\rm DUB} - 3}.
\end{equation}

Regarding the gravitational lensing factor, it is assumed to be the same as in~\cite{Narayan:1996ba}, depending on the binary separation $D^{\rm NAR}$, the angle $\zeta_2^{\rm NAR}$ between the orbital separation and the line of sight and the Einstein radius $R_{E}^{\rm NAR}$. Moreover, for the cases where the neutron star is close or inside the Einstein ring, following~\cite{PhysRev.133.B835}, we pretend it is exactly behind the BH and approximate its image as a ring with angular width equal to the neutron star angular diameter. Therefore, the gravitational lensing factor writes
\begin{equation}
    \left\{
    \begin{array}{ll}
        \frac{(u^{\rm NAR})^2 + 2}{
        u^{\rm NAR}\sqrt{(u^{\rm NAR})^2 + 4}
        } & \text{ for } \|u^{\rm NAR}\| \geq 1 \\
        \min{\left(\frac{(u^{\rm NAR})^2 + 2}{
        u^{\rm NAR}\sqrt{(u^{\rm NAR})^2 + 4}
        }, 2\frac{\sqrt{\frac{4Gm_{\rm BH}^{\rm NAR}D^{\rm NAR}}{c^2}}}{r_{\rm NS}^{\rm NAR}}\right)} & \text{ for } \|u^{\rm NAR}\| \leq 1
    \end{array},
    \right.
\end{equation}
where $u^{\rm NAR} = D^{\rm NAR}\sin{\zeta_2^{\rm NAR}}(R_{E}^{\rm NAR})^{-1}$, while $m_{\rm BH}^{\rm NAR}$ and $r_{\rm NS}^{\rm NAR}$ are the black hole mass and the neutron star radius.

\bibliography{refs.bib}

\end{document}